\documentclass{aa}  

\usepackage{graphicx}
\usepackage{txfonts}
\usepackage{lipsum}
\usepackage{subcaption}         % necessary for continued figures, example in section 3
\usepackage{lscape}             % to rotate a single page table, example in appendix.
\usepackage{placeins}           % useful with \FloatBarrier, to keep 
\usepackage{amsmath}
\usepackage{multirow}
\usepackage{threeparttable}
\usepackage{hyperref}
\usepackage{lineno}
\usepackage{soul}
\usepackage{xcolor}

\usepackage{natbib}
\bibpunct{(}{)}{;}{a}{}{,} % to follow the A&A style

\newcommand{\kms}{km\,s$^{-1}$}
\newcommand{\gaia}{{\it Gaia}}
                                
\begin{document}

%%%%%%%%%%%%%%%%%%%%%%%%%%%%%%%%%%%%%%%%
% if you use custom commands in your title,
% ensure to check your title when submitting!
%%%%%%%%%%%%%%%%%%%%%%%%%%%%%%%%%%%%%%%%
   \title{Measuring the Milky Way potential with the phase snail shape curves}

   %\subtitle{Subtitle}

%%%%%%%%%%%%%%%%%%%%%%%%%%%%%%%%%%%%%%%%
% Please separate each author with the \and command
%
% Use the \corrauth to provide the corresponding
% author address. It will be automatically inserted as 
% footnote in the PDF output.
%
% Please DO NOT include ORCIDs next to author names.
% Instead, please provide an active address for each coauthor:
% it will be automatically extracted by EDPS editorial system, 
% and co-authors will be be able to authenticate their ORCID.
%
% Only authenticated ORCIDs will be taken into account.
% ORCIDs included here will be removed.
%%%%%%%%%%%%%%%%%%%%%%%%%%%%%%%%%%%%%%%%

   \author{Rui Guo\inst{1,2,3}\corrauth{guorui13@nao.cas.cn}        % use \corrauth for the corresponding author
        \and Zhao-Yu Li\inst{2,3,4}\corrauth{lizy.astro@sjtu.edu.cn}
        \and Juntai Shen\inst{2,3,4}\corrauth{jtshen@sjtu.edu.cn}
        \and Jie Wang\inst{1,5,6}
        \and Shude Mao\inst{7}
        }

   \institute{National Astronomical Observatories, Chinese Academy of Sciences, Beijing 100012, China
   \and Department of Astronomy, School of Physics and Astronomy, Shanghai Jiao Tong University, 800 Dongchuan Road, Shanghai 200240, People's Republic of China
   \and State Key Laboratory of Dark Matter Physics, School of Physics and Astronomy, Shanghai Jiao Tong University, Shanghai 200240, People's Republic of China
   \and Key Laboratory for Particle Astrophysics and Cosmology (MOE) / Shanghai Key Laboratory for Particle Physics and Cosmology, Shanghai 200240, People’s Republic of China
   \and Institute for Frontiers in Astronomy and Astrophysics, Beijing Normal University, Beijing 102206, China
   \and University of Chinese Academy of Sciences, 19 A Yuquan Rd, Shijingshan District, Beijing, 100049, China
   \and Deparatment of Astronomy, Westlake University, Hangzhou 310030, Zhejiang Province, China}

   \date{Received March 27, 2026}

% \abstract{}{}{}{}{}
% 5 {} token are mandatory
 %\LEt{***General notes: (A.) I have edited to US English spelling and grammar conventions. (B.) A\&A uses the past tense to describe specific methods used in a paper and the present tense to describe general methods as well as findings, including the findings of recent papers (within the past ten or so years). Kindly make any necessary changes (I have made some, but my edits are by no means exhaustive in this respect). See Sect. 6 of the language guide https://www.aanda.org/for-authors/language-editing/6-verb-tenses}
  \abstract
   {}
   {Phase spirals found in the vertical phase-space ($z-V_z$) indicate that the Galactic disk is in disequilibrium. Although the origin remains elusive, the wrapping of phase snails is dominated by the vertical potential profiles, naturally encoding potential information in their shapes. We attempt to measure the Galactic potential utilizing the snail shape curves.} 
   {To measure the Galactic potential, we extracted the phase snail shape curves from density contrast maps and transformed them into the vertical frequency-action angle ($\Omega_z - \theta_z$) space via orbital integration, then performed linear fitting to derive potential parameters. We verified our method with a test particle simulation with the disk perturbation included within the impulse approximation.}
   {Our method performs well in the test particle simulation, with the recovered Galactic vertical potential profiles in excellent agreement with the input potential models. We also apply our method to the observed phase snails with guiding center radii in the range of $8\le R_g\le11$ kpc. Utilizing snail shape curves only, the model derived vertical potential profiles agree with popular Milky Way potential models. The model yields a Galactic virial mass of $M_{200}= 0.81^{+0.22}_{-0.23} \times10^{12}\, {\rm M_{\odot}}$ and a local dark matter density of $\rho_{dm,\odot}= 0.0076^{+0.0019}_{-0.0021} \,{\rm M_{\odot}\,pc^{-3}}$, consistent with the literature, but with a flatter rotation curve than the observation. To better constrain the Galactic potential, we further combine the phase snails and the rotation curve. The improved model results in $M_{200}= 0.87^{+0.31}_{-0.33} \times10^{12}\, {\rm M_{\odot}}$ and $\rho_{dm,\odot}= 0.0097^{+0.0002}_{-0.0009} \,{\rm M_{\odot}\,pc^{-3}}$. The high precision of $\rho_{dm,\odot}$ results from the joint constraints in the solar neighborhood. We also obtain an estimation of the perturbation time in a range of 400-750 Myr with a median of 564 Myr, consistent with recent measurements.}
   {Our work demonstrates the wealth of information contained in the phase snails to constrain the Galactic potential and mass distribution. We expect to apply our method to the widely existed phase snails across the disk to explore the 3D potential maps and mass distributions in the future.}

   \keywords{The Galaxy --
                Milky Way dynamics --
                Galaxy structure
               }

   \maketitle

   \nolinenumbers

%%%%%%%%%%%%%%%%%%%%%%%%%%%%%%%%%%%%%%%%%%%%%%%%%%%%%%%%%%%%%%
\section{Introduction} 
\label{sec:intro}
The vertical phase-space spiral (snail), a coherent spiral structure in the vertical position-velocity ($z - V_{z}$) plane firstly found by \cite{Antoja2018} utilizing the unprecedented \textit{Gaia} Data Release 2 (DR2; \citealt{Gaia2018}), is a direct manifestation of a dynamical disequilibrium of Milky Way disk. The spiral feature arises from the incomplete phase mixing after a Galactic disk perturbation, which could be due to various mechanisms, such as the interaction with a satellite galaxy (especially the recent pericentric passage of the Sagittarius dwarf; e.g., \citealt{Binney2018, Bland-Hawthorn2019, Laporte2019, Bland-Hawthorn2021, Hunt2021}), the buckling instability due to the central bar \citep{Khoperskov2019}, the dark matter (DM) wake induced after a satellite passage \citep{Laporte2019, Grand2023}, the cumulative effect of smaller intersections with subhalos \citep{Tremaine2023}, the scattering by a clumpy, star-forming interstellar medium \citep{Tepper-Garcia2025}, or the gas inflow along the warp \citep{WangSY2026}.

The phase snail is found widely across the Galactic disk \citep[e.g.,][]{XuY2020, LiZY2021, Hunt2022, Antoja2023}, with the shape varying with the stellar age, chemistry, orbital properties, and spatial location (radius and azimuth) in the Galactic disk \citep[e.g.,][]{TianHJ2018, Bland-Hawthorn2019, Laporte2019, WangC2019, LiZY2020, XuY2020, LiZY2021, Hunt2022, Alinder2023, Antoja2023, Darragh-Ford2023}. Though phase snails usually manifest in the maps encoded by the stellar density, radial and azimuthal velocities \citep{Antoja2018}, they are recently found in maps colored by the residual stellar chemical abundances, mass, and age, i.e., the “Iron Snails” \citep{Frankel2024}, and also in the angular momentum space \citep{Khanna2019, Yaaqib2025}.

Recent studies on the phase snail have focused on several key aspects. First, the time since the disk perturbation ($T_{evol}$) is usually estimated to be between 300 Myr and 900 Myr, through various methods \citep[e.g.,][]{Antoja2018, Binney2018, Bland-Hawthorn2019, LiZY2020, LiHC2021, Antoja2023}, especially by measuring the linear slope of the phase snail in the frequency-action angle space ($\Omega_{z} - \theta_{z}$) \citep[e.g.,][]{LiHC2021, Darragh-Ford2023, Frankel2023}. The perturbation time is found to vary with the angular momentum, $L_{z}$ (or equally the guiding center radius, $R_{g}$), and the azimuthal action angle, $\theta_{\phi}$, showing some structures \citep{Darragh-Ford2023, Frankel2023, Frankel2024, Widmark2026}. Secondly, the amplitude ($\alpha$) of the phase snail, i.e., the relative fraction of stars in the snail feature to the smooth background, also varies with the angular momentum and azimuthal action angle, revealing peaks at several radii \citep{Alinder2023, Frankel2023, Frankel2024}. Both measurements might provide clues for the perturbation mechanism, such as restricting the orbital properties of the interacting satellite. Third, some studies focus on the spatial variation in the snail shape \citep[e.g.,][]{LiZY2021, Hunt2022, Antoja2023} and orientation (the phase angle at the origin or a certain vertical frequency; \citealt{Hunt2021, Hunt2022, Alinder2023, Darragh-Ford2023, Widmark2026}). The snail orientation shows a global trend across the disk that could correspond to the corrugation wave on the disk \citep{Hunt2021, Alinder2023}. Another interesting feature of the phase snail is the two-armed spiral found firstly in the outer disk in the simulation and subsequently in the inner disk of the observation \citep{Hunt2021, Hunt2022}, which can be produced by multiple perturbations \citep{Hunt2022, LinJX2025}, a decelerating bar \citep{LiCD2023}, or a vertical resonance that breaks the disk adiabaticity \citep{Chiba2025}.

The disequilibrium Galactic disk revealed by the phase snail raises questions on the equilibrium assumption of the Galactic disk in previous dynamical studies, as do the north-south asymmetries of the Galactic disk \citep[e.g.,][]{Widrow2012, Bennett2019, GuoR2020}. However, the phase snail shape is determined by the vertical potential, which naturally contains wealthy information about the Galactic mass distribution. Several studies have attempted to constrain the vertical potential combining the phase snail. For example, \cite{LiHC2021, LiHC2023} measure the vertical potential by modeling the phase-space distribution function of giant star samples under equilibrium state, while the phase snail emerges in the residual map after subtracting the best-fit model from the observation. \cite{GuoR2022} fit the north-south asymmetric velocity dispersion profile utilizing a geometric snail model superimposed on a smooth background. While quantitatively explaining several north-south symmetries, they also obtain the vertical potential measurement through Jeans modeling of the background. Moreover, in a series of studies, \cite{Widmark2021a, Widmark2021b, Widmark2022a, Widmark2022b} model the bulk phase-space distribution by the Gaussian mixture model and subtracted it from the data. By fitting the two-dimensional (2D) phase-space residuals with analytical phase snail models constructed through the 1D (vertical) phase-space orbital integration, they determine the vertical potential embedded in the phase-space orbital integration. They obtain robust vertical potential at $z\sim 0.4-0.5$ kpc and utilize these measurements to study the disk properties and the DM disk \citep{Widmark2021b, Widmark2022a}.

In our recent work of \cite{GuoR2024}, we utilize the intersections between the phase snail and $z/V_{z}$ axes, which have either measurable maximum vertical heights ($Z_{max}$) or maximum vertical velocities ($V_{z,max}$), and propose an interpolation method of deriving accurate ($Z_{max}$, $V_{z,max}$) values for these intersections. The relation of $(Z_{max},\ \frac{1}{2}V_{z,max}^{2})$ provides a direct measurement of the vertical potential at these intersections, which is model-independent with no assumptions about the snail shape or mass model. Rather than the intersections in \cite{GuoR2024}, in this work we attempt to utilize the whole snail shape curve and the 1D phase-space orbital integration method to measure the vertical potential, similar to \cite{Widmark2021a}. Unlike them, we first extract the snail shape curves from the observation, and then transform the $z - V_{z}$ phase-space snail into data points in the frequency-action angle space ($\Omega_{z} - \theta_{z}$). We then apply the Markov Chain Monte Carlo (MCMC) technique to compare the linear fitting in $\Omega_{z} - \theta_{z}$ space of different potential models, to obtain parameter estimations of the vertical potential. Though this method now requires mass modeling, the more information provided by the snail shape curves would offer stronger constraints on the vertical potential profile and mass distribution.

The paper is organized as follows. We describe and verify our method with a test particle simulation in Sect. \ref{sec:method}. The observational data used and its model setup are described in Sect. \ref{sec:data}, and the results of the application of our method to the observation are shown in Sect. \ref{sec:results}. The comparisons with relevant works and some discussions are shown in Sect. \ref{sec:discuss}, with the conclusions in Sect. \ref{sec:con}.

%%%%%%%%%%%%%%%%%%%%%%%%%%%%%%%%%%%%%%%%%%%%%%%%%%%%%%%%%%%%%%
\section{Method and simulation verification} 
\label{sec:method}
%%%%%%%%%%%%%%%%%%%%%%%%%%%%%%%%%%%%%%%%%
%%%%%%%%%%%%%%%%%%%%%%%%%%%%%%%%%%%%%%%%%
\subsection{Method description}
\label{ssec:method_desc}
%%%%%%%%%%%%%%%%%%%%%%%%%%%%%%%%%%%%%%%%%
The phase snail arises from the incomplete phase mixing in the vertical phase-space induced by an internal or external perturbation. After the perturbation, ideally, stars are distributed to a similar initial phase angle. As stars with different vertical energies have different vertical orbital frequencies, they gradually wind up into a spiral feature due to the differential rotation in the phase space, which corresponds to a linear relationship in the vertical frequency, $\Omega_{z}$, and vertical phase angle, $\theta_{z}$, space. This linear relationship between $\Omega_{z}$ and $\theta_{z}$ can be utilized to measure the vertical potential. In this work, we attempt to utilize the 1D orbital integration method to transform the snail shape curve extracted from $z - V_{z}$ space to data points in $\Omega_{z} - \theta_{z}$ space, where we can perform linear fitting and apply MCMC technique to obtain the estimation of the vertical potential. The wrapping of the snail and the 1D orbital integration method have been described in great detail in Sects. 2.1 and 2.2 of \cite{Widmark2021a}, which are briefly summarized below.

The 1D orbital integration method is based on the assumption that the vertical and in-plane motions are separable. Thus, the vertical energy\footnote{Strictly speaking, energy $E_{z}$, as well as the potential and kinetic energy, is the specific energy per unit mass. Throughout this paper, we refer to any `specific energy' as simply `energy' for shorthand.}, $E_{z}$, is conserved for a vertically oscillating star, which moves in the vertical phase space following a certain closed orbit. The vertical energy is given by the vertical gravitational potential and vertical kinetic energy according to 
%%%%%%%%%%%%%%%
\begin{equation}
\label{eq:Ez}
    E_{z}= \frac{1}{2} V_{z}^{2}\, +\, \Psi(z) \ ,
\end{equation}
%%%%%%%%%%%%%%%
where the vertical potential energy, $\Psi(z)$, is a reduced energy under a Milky Way potential, $\Phi(R,z)$, as
%%%%%%%%%%%%%%%
\begin{equation}
\label{eq:Phiz}
    \Psi(z)= \Phi(R,z)\, -\, \Phi(R,0)\ .
\end{equation}
%%%%%%%%%%%%%%%

Under a certain vertical potential profile, $\Psi(z)$, the phase-space trajectory of a star is fully determined by its vertical energy, $E_{z}$. The ellipse track of an oscillating star in $z - V_{z}$ space intersects with $z/V_{z}$ axes at $(\pm Z_{max}, 0)$ and $(0, \pm V_{z,max})$, where $Z_{max}$ and $V_{z,max}$ are the maximum vertical height and maximum vertical velocity, respectively. The orbital period, $P(E_{z})$, of the star can be calculated through the 1D orbital integration as 
%%%%%%%%%%%%%%%
\begin{equation}
\label{eq:Pz}
    P\left(E_{z}\right)= \oint \frac{d z}{\left| V_{z} \right|}= 4\int_0^{Z_{max}} \frac{d z}{\sqrt{2\left[ E_{z}-\Psi(z) \right]}} \ ,
\end{equation}
%%%%%%%%%%%%%%%
where the factor of 4 on the right-hand side results from the symmetry in four quadrants of the phase space. The vertical frequency is therefore defined as $\Omega_{z}= 2\pi / P\left(E_{z}\right)$. Both $P(E_{z})$ and $\Omega_{z}$ are functions of $E_{z}$.

The orbital frequency, $\Omega_{z}$ (equally the period $P(E_{z})$), can be calculated through the numerical integration to a relative accuracy of $10^{-6}$. Similarly, we can adopt the numerical integration to calculate the phase angle, $\theta_{z}$, which is defined as the fraction of elapsed time relative to the orbit period. For a given phase-space point $(z, V_{z})$, we first calculated the elapsed time, $t$, of moving from the starting point $(0, V_{z,max})$ to $(|z|, |V_{z}|)$ as
%%%%%%%%%%%%%%%
\begin{equation}
\label{eq:time}
    t = \int_0^{|z|} \frac{d z'}{\sqrt{2\left[ E_{z}-\Psi(z') \right]}} \ .
\end{equation}
%%%%%%%%%%%%%%%
Therefore, according to the orbital symmetry in the four quadrants, the phase angle can be calculated as 
%%%%%%%%%%%%%%%
\begin{equation}
\label{eq:thetaz}
    \theta_{z}(z, V_{z})=
       \begin{cases}
       2\pi t / P & (z\geq0, \, V_{z}\geq0), \\
       \pi - 2\pi t/P & (z\geq0, \, V_{z}<0), \\
       \pi + 2\pi t/P & (z<0, \, V_{z}\leq0), \\
       2\pi - 2\pi t/P & (z<0, \, V_{z}>0). 
       \end{cases}
\end{equation}
%%%%%%%%%%%%%%%
This definition results in the critical linear relation between the phase angle, $\theta_{z}$, and the orbit frequency, $\Omega_{z}$, as given by 
%%%%%%%%%%%%%%%
\begin{equation}
\label{eq:tzOz}
    \theta_{z}= \theta_{z,0} + \Omega_{z} \times T_{evol} ,
\end{equation}
%%%%%%%%%%%%%%%
where $T_{evol}$ is the evolution time since the perturbation and $\theta_{z,0}$ is the initial phase angle.

Assuming a certain vertical potential, the $z-V_{z}$ phase-space snail was transformed to data points in $\Omega_{z} - \theta_{z}$ space through the numerical integration of Eq. \ref{eq:thetaz}. We then performed linear fitting for $\Omega_{z} - \theta_{z}$ space points according to Eq. \ref{eq:tzOz}. After setting up the potential function, we utilized the MCMC technique to obtain the best-fit parameters of the potential with the {\tt EMCEE} package \citep{Foreman-Mackey2013}. The resultant posterior parameter distributions were compared to the literature, and the obtained vertical potential was also compared to popular Milky Way potential models.

%%%%%%%%%%%%%%%%%%%%%%%%%%%%%%%%%%%%%%%%%
\subsection{Method verification with test particle simulation}
\label{ssec:method_test}
%%%%%%%%%%%%%%%%%%%%%%%%%%%%%%%%%%%%%%%%%
%%%%%%%%%%%%%%%%%%%%%%%%%%%%%%%%%%%%%%%%%
\subsubsection{Test particle simulation}
\label{sssec:simu_set}
%%%%%%%%%%%%%%%%%%%%%%%%%%%%%%%%%%%%%%%%%
We verified our method with a test particle simulation similar to our previous works \citep{LiZY2021, GuoR2024}, with the initial condition sampling and 3D orbital integration performed by {\tt AGAMA} \citep{Vasiliev2019}. The input Milky Way potential adopted was Model I from \cite{Irrgang2013} (hereafter Irrgang13I). The test particles traced a thin disk population with a \texttt{QuasiIsothermal} distribution function (DF; \citealt{Binney2010, Binney2011}) and basic properties (scale lengths and velocity dispersion) set to be dynamically cold. The disk perturbation was set to mimic the impulse approximation by imposing a downward vertical velocity with no vertical displacement. We added a vertical velocity kick following a Gaussian distribution of $-30\pm5$ {\kms} to all particles and evolved the simulation for 500 Myr. This snapshot was used for the method verification here. Therefore, the initial phase angle was $\theta_{z,0}= \pi$ and the evolution time of the perturbation was $T_{evol}= 500$ Myr.

The potential of Irrgang13I consists of a Plummer bulge \citep{Plummer1911}, a Miyamoto-Nagai disk \citep{MN1975}, and a specific DM halo profile constructed to have approximately flat rotation curve and finite halo mass at a cutoff radius. The corresponding mass and potential functions of different components are described in detail in \cite{Irrgang2013} and our Eqs. \ref{eq:bulgeP}-\ref{eq:IDMM}. The parameters of this potential are $\mathbf{p}= (M_b,\ M_d,\ M_h,\ b_b,\ a_d,\ b_d,\ a_h,\ \Lambda)$, where $M_b,\ M_d,\ M_h$ are the scale masses of the bulge, disk, and DM halo, respectively. $b_b$ is the scale length of the bulge. $a_d$ and $b_d$ are the scale lengths of the disk. $a_h$ and $\Lambda$ are the scale radius and cutoff radius of the halo profile. The input potential parameters adopted in our test particle simulation are the best-fit values taken from \cite{Irrgang2013}.

%%%%%%%%%%%%%%%%%%%%%%%%%%%%%%%%%%%%%%%%%
\subsubsection{Snail shape measurement}
\label{sssec:simu_snail_shape}
%%%%%%%%%%%%%%%%%%%%%%%%%%%%%%%%%%%%%%%%%
We regard the ridge line of the phase snail in the density contrast map as the snail shape curve. This definition is different from that used in \cite{Antoja2023}, who adopted the pixels with maximum gradients as the snail shape based on an edge-detector algorithm. We utilized a similar method as \cite{LiZY2021} to extract the shape curve of the phase snail. Our measurement of the snail shape curve is also similar to the procedure of ridge line extraction in \cite{Darragh-Ford2023}, different in the space performed ($z-V_{z}$ space vs. $J_{z}-\theta_{z}$ space).

Our snail shape extraction was performed as following three steps. First, we derived the phase-space density contrast map. The data were separated into different radial bins according to the guiding center radius, $R_{g}$, with a bin width of 0.6 kpc. The $R_{g}$-binned snails show a higher clarity than the galactocentric radius binned ones, and thus can reduce the uncertainty in the snail shape measurement. The 2D density ($N$) map with a pixel size of 0.02 kpc $\times$ 1 $\rm km\,s^{-1}$ was convolved with a Gaussian kernel ($\sigma= 4$ pixels) to derive a smooth background ($\widetilde{N}$), which was then subtracted to obtain the 2D density contrast map ($\Delta N= N/\widetilde{N} -1$). We further smoothed the density contrast map with a Gaussian kernel ($\sigma= 2.1$ pixels) to enhance the snail signal.

\begin{figure}[bth!]
\centering
\includegraphics[width=0.85\columnwidth]{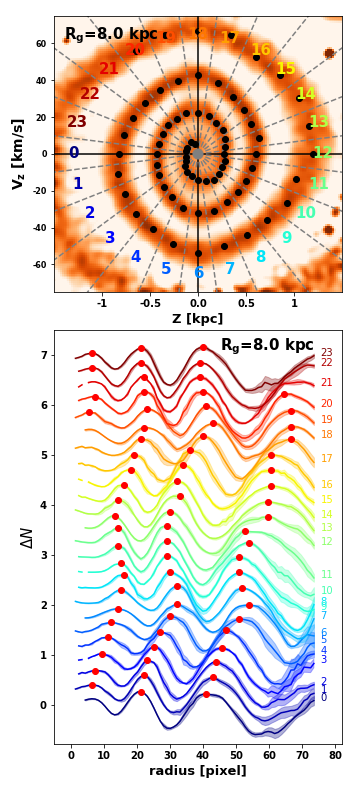}
\caption{
\label{fig:find_peaks}
 \textit{Top}: Illustration of wedge decomposition at $R_{g}= 8.0$ kpc. The orange map shows the density contrast map. The 24 wedges are separated by the dashed gray lines and indicated by the colored numbers, representing a counterclockwise arrangement of wedges in $z-V_{z}$ phase-space starting from $z<0$ axis from small (blue) to large (red) numbers. \textit{Bottom}: Density contrast profiles for different wedges of the phase snail at $R_{g}= 8.0$ kpc. The colored lines represent the median density contrast profiles for all pixels inside the corresponding wedges, with shaded regions indicating the uncertainties. The vertical shifts are applied for visual clarity. The red dots label the snail shape curve (the ridge line) found by the peak-finding program, overplotted in the upper panel by black points.}
\end{figure}

Second, we separated the density contrast map into 24 wedges ($15^{\circ}$ each) in $z-V_z$ space. As illustrated in the upper panel of Fig. \ref{fig:find_peaks}, the wedges are separated by dashed gray lines and indicated by colored numbers, arranged in a counterclockwise order in phase-space starting from $z<0$ axis from small (blue) to large (red) numbers. For each wedge, we plot the residual density, $\Delta N$, as a function of phase-space radius (in pixels) for all pixels inside. We then take the median profile as the 1D density contrast profile, with the uncertainty taken from the standard deviation. The density contrast profiles of 24 wedges are shown as the corresponding colored lines in the lower panel of Fig. \ref{fig:find_peaks}. They usually show several peaks and troughs with radial shift across the wedges, representing the different wraps of the phase snail. Finally, we utilize the implementation {\tt signal.find\_peaks} from \texttt{scipy} to find the density peaks, shown as the red dots in the lower panel of Fig. \ref{fig:find_peaks} and overplotted in the upper panel by the black points. We connect these points to represent the snail shape curve in $z-V_z$ space. We ignore the outmost region showing obvious large uncertainties and some inner peaks with small peak amplitudes.

\begin{figure*}[htbp]
\centering
\includegraphics[width=0.85\textwidth]{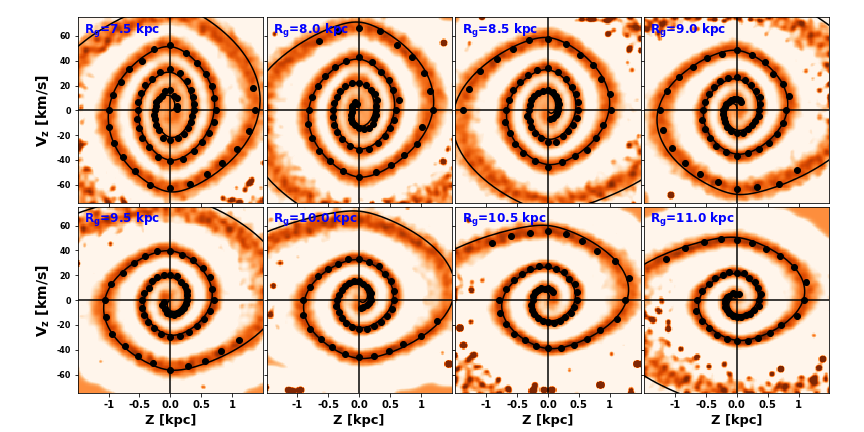}
\caption{
\label{fig:simu_overplot}Phase snails of the test-particle simulation at different guiding center radii from 7.5 to 11.0 kpc. The measured snail shapes (black points), as well as the toy model snails calculated from the 1D orbit integration (black lines), are overplotted on the density contrast (orange) maps.}
\end{figure*}

In Fig. \ref{fig:simu_overplot}, the measured snail shape curves for 8 $R_{g}-$binned snails (with bin center ranging from 7.5 kpc to 11 kpc) are superimposed on the density contrast $\Delta N$ maps, as shown by the black points. The black lines in Fig. \ref{fig:simu_overplot} are the simple toy model snails calculated through the 1D orbital integration under the input vertical potential. The toy model snails are initially located at the negative $V_{z}$ axis, mimicking an initial phase angle perturbation of $\theta_{z,0}= \pi$. We calculated their orbital frequencies and phase angles at $t=500$ Myr, and then transformed these $\Omega_{z}-\theta_{z}$ space points into the toy model snail displayed in $z-V_{z}$ space. As shown in Fig. \ref{fig:simu_overplot}, there is a good consistency between the density contrast maps, the measured snail shape curves and the toy model snails, qualitatively indicating a good representation of the phase snail by our shape detection and the separability of the in-plane and vertical motions at $z \lesssim 1.5$ kpc. In outer region ($z > 1.5$ kpc), the toy model snails deviate from the density contrast maps, which is due to the coupling of in-plane and vertical motions at large vertical heights. Thus, we only focus on the inner region where also shows smaller uncertainty in the observation.

For an additional quantitative test of the separability assumption of 1D orbital integration, we compared the snail shapes and vertical energy profiles under 1D and 3D integrations. For the 3D integration, we integrated initial snails in a 3D Milky Way potential, i.e., the Irrgang13I potential adopted in our test particle simulation. The initial conditions of the snails are $( R=R_{g},\ \phi=0,\ z=0,\ V_{R}= 0,\ V_{\phi}= V_{c}(R_{g}),\ V_{z}= -V_{z,max} )$ in cylindrical coordinates. These orbits have initial azimuthal velocities equal to the circular velocities with zero radial actions, which are then integrated for 500 Myr. The evolved snails are similar to the black points in Fig. \ref{fig:simu_overplot}, showing good consistency with the black lines at $|z| < 1.5$ kpc, i.e., the snails ignoring the in-plane motions under 1D orbital integration. To quantify the difference between the 1D and 3D integrations, we calculated the vertical potentials represented by the snails under these two integrations. The potential difference profiles display wave-like patterns, and show slightly large discrepancy at points close to the mid-plane points ($V_z \sim \pm V_{z,max}$), which are more easily influenced by the in-plane motions. Nevertheless, we find that the median difference of these two potential profiles is usually smaller than 2\%, and the maximum difference at $|z| < 1.5$ kpc is smaller than 5\%.

%%%%%%%%%%%%%%%%%%%%%%%%%%%%%%%%%%%%%%%%%
%%%%%%%%%%%%%%%%%%%%%%%%%%%%%%%%%%%%%%%%%

%%%%%%%%%%%%%%%%%%%%%%%%%%%%%%%%%%%%%%%%%
%%%%%%%%%%%%%%%%%%%%%%%%%%%%%%%%%%%%%%%%%

Note that, according to \cite{GuoR2024}, the ridge line of a $R_{g}-$binned snail is located slightly inside that of the $R$-binned snail, indicating a shallower vertical potential with a slightly larger potential radius. This effect is similar to the asymmetric drift effect, i.e., there are more stars at $R$ actually lagging the local standard of rest with smaller $R_g$. We proposed an empirical method to correct for this effect and assign potential radii for $R_{g}$-binned snails \citep{GuoR2024}. For different $R$-bins with a same width, we measure their median $R_g$, building a relation between $R$ and median-$R_g$. We regard $R$ as the potential radius $R_{pot}$ of a $R_g$-binned snail centered at the median-$R_g$. This correction works well for this test particle simulation and shows a difference of about $0.1-0.15$ kpc between the potential radii of $R_{g}-$ and $R-$binned snails (see their Fig. 3). Ignoring this effect will result in a systematic underestimation of the vertical potential. See more details in Sect. 2.2.2 of \cite{GuoR2024}. This effect has been considered in building the toy model snails in Fig. \ref{fig:simu_overplot}, and also in comparing the recovered vertical potential with the input profiles in later figures. We take the potential profiles at the corresponding potential radii of $R_{g}-$binned snails according to the empirical correction as the input profiles, which agrees well with the density contrast maps and model predict potential profiles.

%%%%%%%%%%%%%%%%%%%%%%%%%%%%%%%%%%%%%%%%%
\begin{table}
\centering
\caption{Parameters in recovering the vertical potential with the simulated phase snails.}
\label{tab:paras_Irr}
\begin{tabular}{lrrrr}
\hline
\hline
\textbf{Parameter} & \textbf{Unit} & \textbf{Input} & \textbf{Prior} & \textbf{Posterior} \\
(1) & (2) & (3) & (4) & (5) \\
\hline
\(M_b\) & $10^{10}{\rm M}_{\odot}$ & 0.95 & [0.1, 2.0] & $1.23^{+0.53}_{-0.68}$ \\
\(b_b\) & kpc & 0.23 & [0.1, 0.5] & $0.304^{+0.136}_{-0.136}$ \\
\(M_d\) & $10^{10}{\rm M}_{\odot}$ & 6.64 & [0.5, 20] & $7.22^{+0.36}_{-0.36}$ \\
\(a_d\) & kpc & 4.22 & [1, 8] & $4.126^{+0.313}_{-0.250}$ \\
\(b_d\) & kpc & 0.292 & [0.05, 1] & $0.289^{+0.008}_{-0.007}$ \\
\(M_h\) & $10^{10}{\rm M}_{\odot}$ & 2.367 & [0.2, 100] & $37.75^{+35.45}_{-25.76}$ \\
\(a_h\) & kpc & 2.562 & [1, 50] & $33.298^{+11.432}_{-13.063}$ \\
\(\Lambda\) & kpc & 200 & [100, 250] & $178^{+50}_{-54}$ \\
\(T_{\text{evol}}\) & Myr & 500 & [0, 2000] & $491^{+22}_{-19}$ \\
\(\theta_{z,0}\) & rad & \(\pi\) & [0, 2\(\pi\)] & $4.12^{+0.98}_{-1.21}$ \\
\hline
\end{tabular}
\begin{tablenotes}
\small
\item \textbf{Notes.} Columns are: (1) Parameter names; (2) Units; (3) Input values taken from the Model I of \citet{Irrgang2013}; (4) Priors adopt in the MCMC; (5) Median values and $1\sigma$ uncertainties of 1D marginalized posterior distributions from MCMC.
\end{tablenotes}
\end{table}
%%%%%%%%%%%%%%%%%%%%%%%%%%%%%%%%%%%%%%%%%

%%%%%%%%%%%%%%%%%%%%%%%%%%%%%%%%%%%%%%%%%
\subsubsection{MCMC fitting to the simulated snails}
\label{sssec:simu_MCMC}
%%%%%%%%%%%%%%%%%%%%%%%%%%%%%%%%%%%%%%%%%
We constructed eight phase snails with $R_g$ ranging from 7.5 kpc to 11 kpc to recover the vertical potential using our method. The snail shape curves detected are indicated by black dots in Fig. \ref{fig:simu_overplot}, usually representing two or three snail wraps up to a vertical height of $\sim 1.5$ kpc. The vertical orbital frequencies, $\Omega_z$, and phase angles, $\theta_z$, of these snail shape points were then calculated through 1D numerical integration under a certain vertical potential. As $\theta_z$ is in the range of $[0, 2\pi]$, we manually add $2k\pi$ ($k$ is an integer) to points belonging to different wraps. Thus the points could form a continuous curve in $\Omega_{z} - \theta_{z}$ space. Assuming a certain potential model, we utilized the MCMC technique to linearly fit the curve in $\Omega_{z} - \theta_{z}$ space for different parameter sets and to find the posterior distributions and the best-fit parameters. The input parameters, including the perturbation parameters $T_{evol}$ and $\theta_{z,0}$, and the flat priors adopted are listed in the third and fourth columns of Table \ref{tab:paras_Irr}, respectively.

\begin{figure*}[htbp]
\centering
\includegraphics[width=0.85\textwidth]{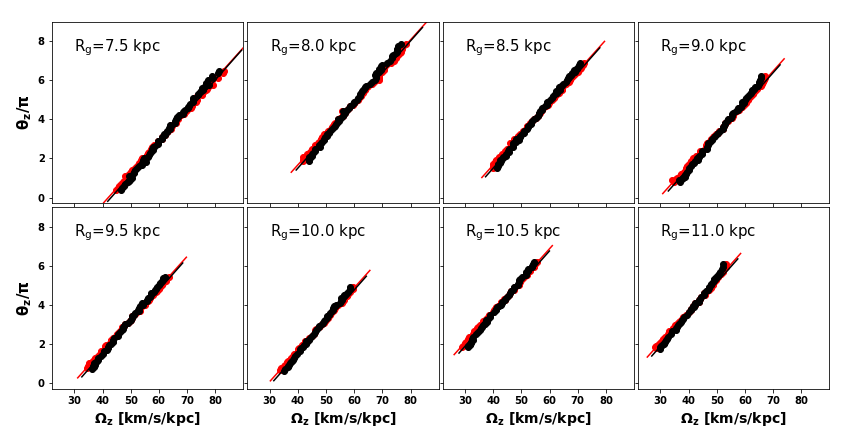}
\caption{
\label{fig:simu_Oz_thetaz}Comparison of the snail shape curves transformed into $\Omega_{z} - \theta_{z}$ space for different $R_g$ bins based on the input potential (Irrgang13I, black points) and the best-fit potential (red points). The solid lines are the linear fitting to the points with the slope indicating the perturbation time $T_{evol}$.}
\end{figure*}

The detected snails transformed into $\Omega_{z} - \theta_{z}$ space under the input vertical potential Irrgang13I and the best-fit potential from MCMC are shown by the black and red dots in Fig. \ref{fig:simu_Oz_thetaz}, respectively. They both have good linear relationships and are highly overlapped in $\Omega_{z} - \theta_{z}$ space, indicating a good consistency between the best-fit and input. The solid lines in Fig. \ref{fig:simu_Oz_thetaz} display the linear fitting to the points, with the slopes indicating the evolution time of perturbation $T_{evol}$. There is also a good recovery of the perturbation time. Note that the points in outer wraps have larger orbital periods, and thus lower frequencies, leading to smaller phase angles. 

\begin{figure*}[htbp]
\centering
\includegraphics[width=0.85\textwidth]{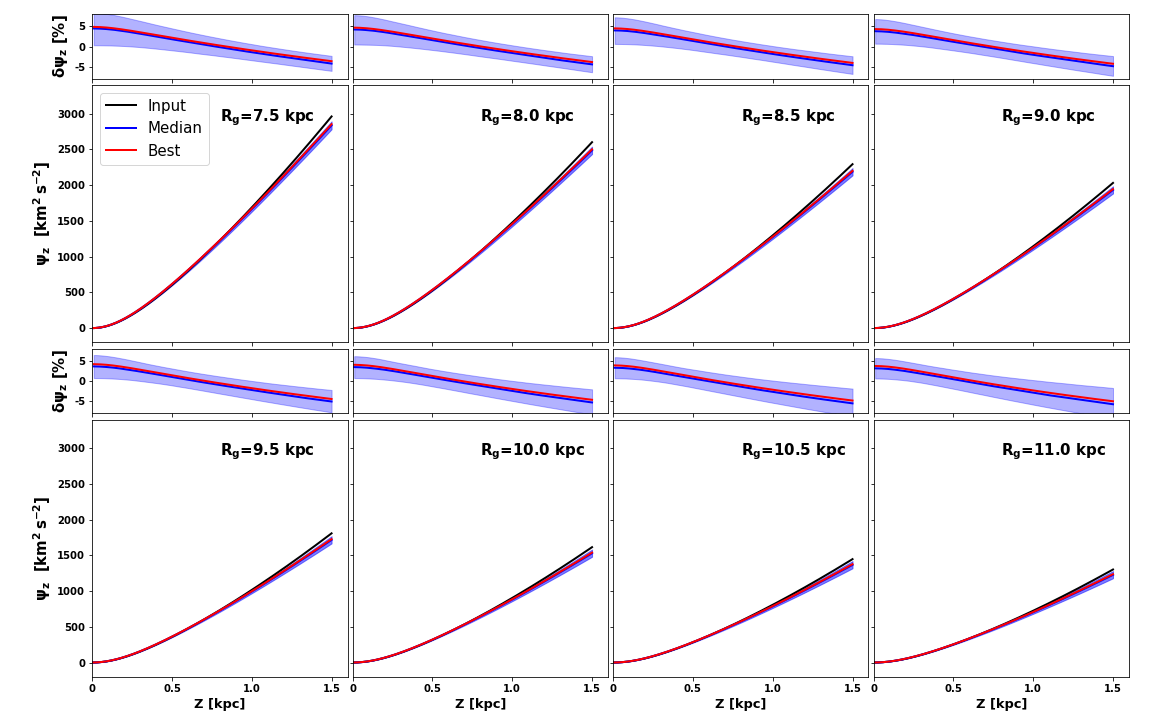}
\caption{
\label{fig:simu_phiz}Comparison of the input vertical potential (black lines) with those from MCMC results. The red lines are the potential profiles from MCMC best-fit parameters, while the blue lines with shaded regions display the median and $1\sigma$ uncertainties of all posterior parameter sets. \textit{Second and fourth rows}: Direct comparison of potential. \textit{First and third rows}: Relative difference (as a percentage) of MCMC predictions relative to the input.}
\end{figure*}

Figure \ref{fig:simu_phiz} shows the comparison of the input vertical potential and those from MCMC results. The potentials from the best-fit parameters and the median of all posterior parameter sets are illustrated by the red and blue lines, respectively, while the input is indicated by the black lines. The second and fourth rows show the direct comparison of potential, while the first and third rows display the relative difference (as a percentage) of MCMC predictions relative to the input. Our recovered vertical potential is in excellent agreement with the input, with a relative difference smaller than 5\% in the region considered, demonstrating a nice performance in measuring the vertical potential with our method.

The residual profiles in Fig. \ref{fig:simu_phiz} show a same biased pattern with overestimation at low $z$ and underestimation at high $z$. This pattern is due to the small discrepancy between the measured snail shapes (the black points) and the 1D toy model snails (the black lines) shown in Fig. \ref{fig:simu_overplot}. The measured snail shapes always locate inside (outside) the toy model snails in the outer (inner) phase space, indicating smaller (larger) vertical energy at high (low) $z$. This indicates a biased potential profile represented by the measured snail shape. It will shift the points with small (large) $\Omega_{z}$ to smaller (larger) $\Omega_{z}$ in $\Omega_{z} - \theta_{z}$ space, as shown by the red points in Fig. \ref{fig:simu_Oz_thetaz}, leading to a flatter slope with a slightly smaller $T_{evol}$. The situation is similar to the arbitrarily scaled potential in Appendix \ref{sec:scaled_pot}, i.e., the blue lines shown in Fig. \ref{fig:scale_up}. This bias in the predicted potential could be attributed to two reasons. First, our method of 1D orbital integration is an approximation to the 3D motions. The measured snail shapes show better consistency with the density contrast maps than the toy model snails, as shown in Fig. \ref{fig:simu_overplot}. The difference between the 1D and 3D integrations, or equally the ignorance of the in-plane motions, result in the difference between the measured snail shape and the 1D toy model snail. Second, the potential radius correction applied to the $R_g$-binned snails may also contribute to the biased potential measurement. This correction is applied both in the potential prediction and the choice of the `input' potential. Therefore, maybe a slightly biased potential radius correction may lead to the discrepancy in the derived potential and thus the slope $T_{evol}$. As shown in Fig. \ref{fig:simu_phiz}, applying slightly shallower input potential profiles with slightly larger potential radii may lead to a better consistency between the model predictions and the input.

\begin{figure*}[htbp]
\centering
\includegraphics[width=0.8\textwidth]{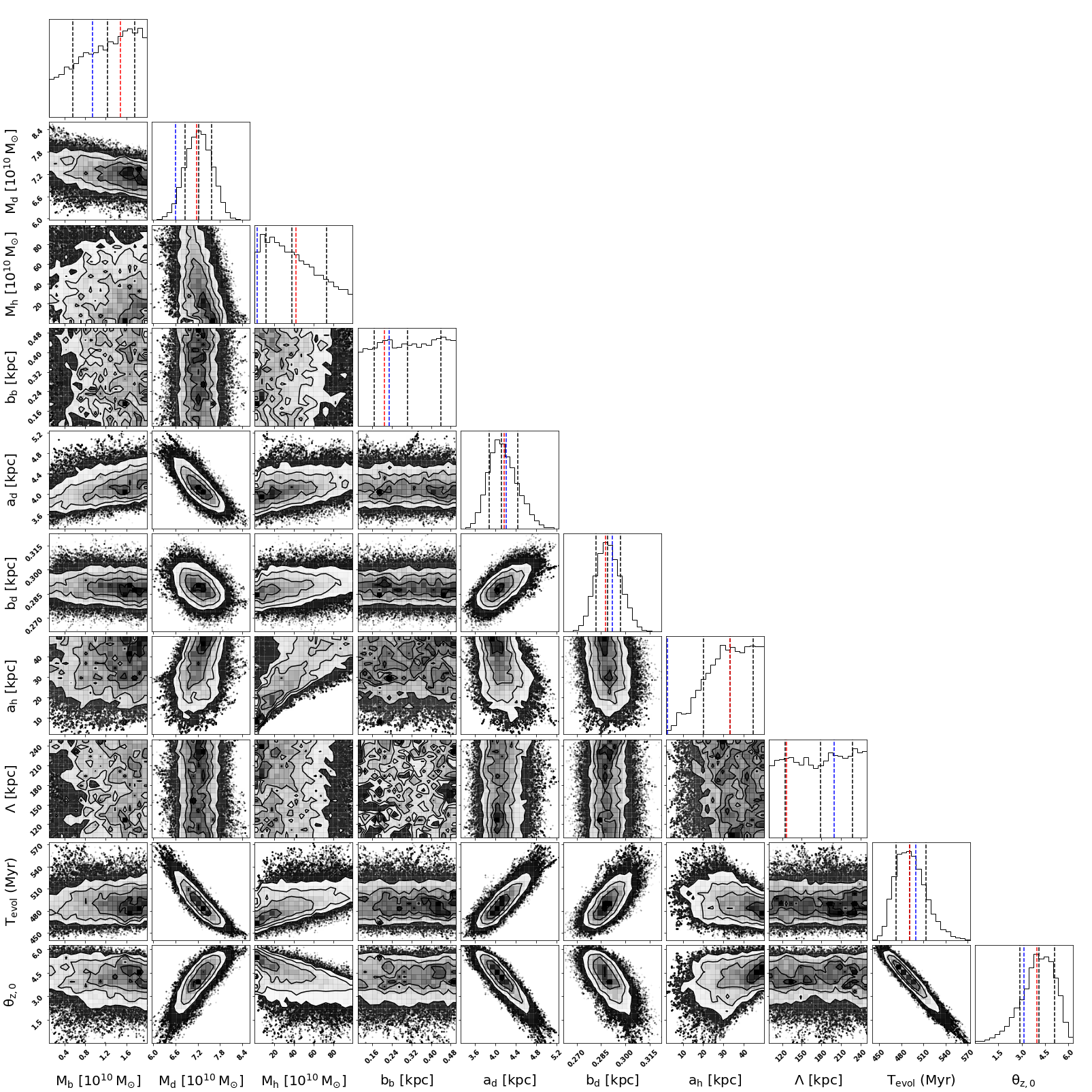}
\caption{
\label{fig:simu_paras}Posterior distributions of MCMC parameters. The histograms are the 1D marginalized distributions of each parameter, with the dashed lines indicating the input (blue), the best fit (red), and the median values and $1\sigma$ uncertainties (black), respectively. The disk parameters are well constrained due to the dominance of the disk component in the vertical potential profile. However, they are significantly degenerate with snail parameters $T_{evol}$ and $\theta_{z,0}$, illustrating the degeneracy between deducing the vertical potential and the perturbation time.}
\end{figure*}

The corner plot of the parameters' posterior distributions is shown in Fig. \ref{fig:simu_paras}.  The median values and $1\sigma$ uncertainties of posterior parameters from MCMC, i.e., the dashed black lines in the histograms of Fig. \ref{fig:simu_paras}, are shown in the fifth column of Table \ref{tab:paras_Irr}. We note that there are poor constraints on the bulge and halo potential, which is due to the minor contributions to the vertical potential profiles at $z< 1.5$ kpc from these two components. There is some degeneracy between the parameters of the disk potential, i.e., $M_d,\ a_d,\ b_d$. Nevertheless, the total vertical potential profile is well recovered as indicated in Fig. \ref{fig:simu_phiz}. We also tried to apply different mass models, such as the single or double exponential disks, to recover the vertical potential utilizing the same snail shape curves. The different mass models converge on the same total vertical potential.

There exists strong degeneracy between the disk parameters and $T_{evol}$ and $\theta_{z,0}$. As shown in Fig. \ref{fig:simu_paras}, there is a negative correlation between $T_{evol}$ and disk mass $M_d$, and positive correlations between $T_{evol}$ and disk scale length $a_d$ and scale height $b_d$. This degeneracy could be understood from the tests in Appendix \ref{sec:scaled_pot}, where we investigate the variation in the curve in $\Omega_{z} - \theta_{z}$ space with the vertical potential deviating from the input one by arbitrarily scaling it. Basically, a globally increasing (steeper) potential will shift the snail to larger $\Omega_z$ in $\Omega_z - \theta_z$ space, with slightly decreased slope, i.e., smaller $T_{evol}$. Decreasing $a_d$ and $b_d$ will lead to steeper vertical potential and thus smaller $T_{evol}$, resulting in the positive correlations between them. The globally scaled potential will also result in a zigzag pattern for the curve in $\Omega_z - \theta_z$ space, leading to a large dispersion in the linear fitting. These globally scaled situations dominate the degeneracy between the disk parameters and $T_{evol}$. For other situations, when the potential of high $z$ and low $z$ regions are scaled oppositely, their shifts in $\Omega_z - \theta_z$ space are also in opposite directions, resulting in significantly changed slope and $T_{evol}$. In some extreme cases, when the potential at high $z$ is scaled up while that at low $z$ is scaled down, the $\Omega_z - \theta_z$ curve show roughly linear relations but with strongly biased slopes $T_{evol}$, up to an overestimation of 5 times. These tests help to understand the uncertainties of the potential recovery and the degeneracy between parameters, especially between the potential parameters and the perturbation time $T_{evol}$.

Overall, our method recovers the input vertical potential profiles well and shows a good linear relation after transforming the phase snail into $\Omega_{z} - \theta_{z}$ space, as shown in Fig. \ref{fig:simu_Oz_thetaz}. The accurate snail shape measurement in the test particle simulation leads to this nearly ideal case. The best-fit and median values of $T_{evol}$ are 492 Myr and $491^{+22}_{-19}$ Myr, respectively, while those values for $\theta_{z,0}$ are $1.27 \pi$ and $1.31^{+0.31}_{-0.38}\, \pi$, respectively. The systematic bias of 10 Myr corresponds to the slightly biased potential predicted by our method, which could be more intrinsically attributed to the ignorance of the in-plane motions in the 1D orbital integration and the potential radius correction for $R_g$-binned snails. In addition, the poor constraint on $\theta_{z,0}$ results from the fact that a small variation in the slope of $T_{evol} \sim 500$ Myr would lead to a large uncertainty in the intercept $\theta_{z,0}$ in the linear fitting.

%%%%%%%%%%%%%%%%%%%%%%%%%%%%%%%%%%%%%%%%%
%%%%%%%%%%%%%%%%%%%%%%%%%%%%%%%%%%%%%%%%%

%%%%%%%%%%%%%%%%%%%%%%%%%%%%%%%%%%%%%%%%%
%%%%%%%%%%%%%%%%%%%%%%%%%%%%%%%%%%%%%%%%%

%%%%%%%%%%%%%%%%%%%%%%%%%%%%%%%%%%%%%%%%%
%%%%%%%%%%%%%%%%%%%%%%%%%%%%%%%%%%%%%%%%%

%%%%%%%%%%%%%%%%%%%%%%%%%%%%%%%%%%%%%%%%%%%%%%%%%%%%%%%%%%%%%%
\section{Data} 
\label{sec:data}
%%%%%%%%%%%%%%%%%%%%%%%%%%%%%%%%%%%%%%%%%%%%%%%%%%%%%%%%%%%%%%
We adopted the standard Galactocentric cylindrical coordinate system ($R,\, \phi,\, z$) for the observation, with $\phi<0$ toward the direction of Galactic rotation and $z$ pointing toward the north Galactic pole. The distance of the Sun to the Galactic center is $R_{0}= 8.34$ kpc \citep{Reid2014}, with a vertical height above the Galactic plane of $z=0.025$ kpc \citep{Juric2008}. For the solar motion, we adopted a radial velocity of $V_{R,\odot}= 11.1$ {\kms} \citep{Schonrich2010}, an azimuthal velocity $V_{\phi,\odot}= -252.2$ {\kms} and a vertical velocity $V_{z,\odot}= 8.0$ {\kms}, calculated from the proper motions of Sgr A$^{*}$ \citep{Reid2004} and the adopted solar distance $R_{0}$.

%%%%%%%%%%%%%%%%%%%%%%%%%%%%%%%%%%%%%%%%%
\subsection{Gaia snails}
\label{ssec:obs_data}
%%%%%%%%%%%%%%%%%%%%%%%%%%%%%%%%%%%%%%%%%
We used the data from {\gaia} DR3 \citep{GaiaDR32023} to extract the snail shape curves at different radii. We selected stars with radial velocities available, i.e., the {\gaia} DR3 RVS which contains about 33.8 million objects. We further applied two quality cuts: (1) parallax or distance cut: the parallax over error $\varpi/\sigma_{\varpi}>5$; (2) astrometric quality cut: the renormalized unit weight error RUWE$<1.4$. The final sample contains 27,209,191 stars. We utilized the geometric distances from \cite{Bailer-Jones2021}.

We calculated the stellar guiding center radii, $R_{g}$, based on the rotation curve from \cite{ZhouY2023} and then separated the data into different $R_{g}$ bins, which show higher clarity of snails than the $R$-binned samples. Note that, as we adopted a slightly larger solar radius, $R_{0}$, than \cite{ZhouY2023} ($R_{0}= 8.122$ kpc), we manually increased the radii and circular velocities of the rotation curve measurements from \cite{ZhouY2023} by 0.218 kpc and 6.6 {\kms}, respectively, to simply compensate for the $R_{0}$ difference. The value of 6.6 {\kms} results from the same Sgr A$^{*}$ proper motions but different solar radii. Thus, the circular velocity at the solar position is now $V_{c,\odot}=240.6$ {\kms} ($V_{c,\odot}=234.04$ {\kms} in \citealt{ZhouY2023}). $R_{g}$ was calculated as the radius of a circular orbit with the same angular momentum ($L_{z}$) as the star, i.e., $R_{g}= -RV_{\phi} \big/ V_{c}(R_{g})$, where $V_{\phi}$ is negative in our coordinates and $V_{c}$ is the circular velocity. We built an empirical $L_{z}-R_{g}$ relation according to the modified rotation curve of \cite{ZhouY2023}, to calculate $R_{g}$ through the interpolation of the $L_{z}-R_{g}$ relation based on $L_{z}$.

\begin{figure}[bth]
\centering
\includegraphics[width=0.9\columnwidth]{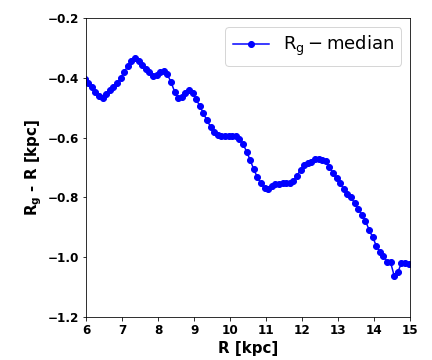}
\caption{
\label{fig:obs_Rpot}Difference between a radial bin center and its $R_{g}$ for the observation. $R_{g}$ was calculated according to the shifted rotation curve from \cite{ZhouY2023}. This relation was utilized to relate a $R_{g}$-binned snail to the potential radius $R_{pot}$.}
\end{figure}

\cite{GuoR2024} point out that the derived vertical potential profile at $R_g$ is slightly shallower than the real vertical potential profile at the same Galactocentric radius $R$ ($= R_g$). The profile measured at $R_g$ actually corresponds to that at a slightly larger radius, which is called the potential radius $R_{pot}$. We utilized the empirical method mentioned in the previous section to correct for this effect and relate the potential radius $R_{pot}$ to the $R_{g}$-binned snail. The empirical relation between $R_{g}$ and the potential radius $R_{pot}$ for the observational sample is shown in Fig. \ref{fig:obs_Rpot}. For the radial range of our sample ($7<R_g<11$ kpc), $R_g$ is about 0.4–0.8 kpc smaller than $R_{pot}$. Note that this relation is empirical and changes with the sample selection and also the rotation curve adopted. We also tried another empirical correction based on a different rotation curve of \cite{HuangY2016}, which shows significantly different $R_{pot}$ correction at $R_g >10$ kpc. Nevertheless, we obtain similar constraints on the model parameters, as the model is dominated by the constraint at inner radii where have more data points and a higher accuracy. Moreover, ignoring this $R_{pot}$ correction would result in a significant underestimation of the vertical potential, especially at inner radii. 

%%%%%%%%%%%%%%%%%%%%%%%%%%%%%%%%%%%%%%%%%
%%%%%%%%%%%%%%%%%%%%%%%%%%%%%%%%%%%%%%%%%

\begin{figure*}[htbp]
\centering
\includegraphics[width=0.85\textwidth]{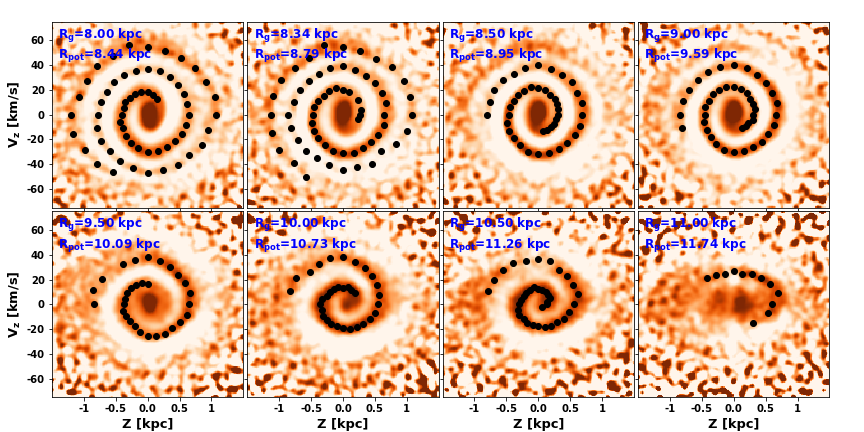}
\caption{\label{fig:obs_overplot}Same as Fig. \ref{fig:simu_overplot} but for snail shape measurements for the observed phase snails. The pixel size is 0.02 kpc $\times$ 1 {\kms}. The $R_{g}$ bin centers and the corresponding potential radii $R_{pot}$ are labeled on the upper left corner of each panel.}
\end{figure*}

We measured the snail shape curves at eight radial bins with centers at $R_{g}= 8, 8.34, 8.5, 9, 9.5, 10, 10.5, 11$ kpc, with a bin width of $\Delta R_{g}= 0.5$ kpc. Phase snails within 8 kpc are not explored due to the complexity introduced by the two-armed phase spirals or secondary branches \citep{Hunt2022, Antoja2023}. On the other hand, at $R_g > 11$ kpc, the phase snails are hardly discernible. We trace the snails in a phase space region of $|z|$< 1.5 kpc and $|V_{z}|< 75$ {\kms}, with a pixel size of 0.02 kpc $\times$ 1 {\kms}. The measured snail shape curves are indicated by the black dots in Fig. \ref{fig:obs_overplot}, superimposed on the density contrast maps at different $R_g$. The snails are well represented by a large number of points (321 points in total) with $1-3$ wraps, which are then utilized to measure the vertical potential of our Milky Way.

%%%%%%%%%%%%%%%%%%%%%%%%%%%%%%%%%%%%%%%%%
%%%%%%%%%%%%%%%%%%%%%%%%%%%%%%%%%%%%%%%%%

%%%%%%%%%%%%%%%%%%%%%%%%%%%%%%%%%%%%%%%%%
\subsection{Mass model setup for the observational data}
\label{ssec:obs_model}
%%%%%%%%%%%%%%%%%%%%%%%%%%%%%%%%%%%%%%%%%
When measuring the vertical potential with the observed snail shape curves, we utilized a potential model consisting of a Plummer sphere bulge \citep{Plummer1911}, a Miyamoto-Nagai disk \citep{MN1975}, and a NFW DM halo \citep{NFW1996}. The mass and potential profiles, and the related rotation curves, are presented in Appendix \ref{sec:app_mfunc}. The parameters of this potential model are $\mathbf{p}= (M_b,\ M_d,\ M_h,\ b_b,\ a_d,\ b_d,\ a_h)$, where $M_b,\ M_d,\ M_h$ are the scale masses of the bulge, disk, and DM halo, respectively. $b_b$ is the scale length of the bulge. $a_d$ and $b_d$ are the scale lengths of the disk. $a_h$ is the scale radius of the halo profile. This mass model replaces the halo profile of the test particle simulation with a NFW profile, without a cutoff radius. The NFW profile is chosen for more convenient comparisons with other works and is more suitable to the observed declining rotation curve. In addition, we have tested to apply different mass models to the same snail shape curves in the test particle simulation. The different mass models converge on the same total vertical potential encoded in the snail shapes. As the bulge has less influence on the radial range explored ($8 \le R_{g} \le 11$ kpc), we fix the bulge parameters as $M_b= 1.067 \times 10^{10}$ $\rm M_{\odot}$ and $b_d= 0.3$ kpc, directly adopted from \cite{Pouliasis2017}. We also apply a broad Gaussian prior on the disk mass $M_d$, which is $5.17\pm1.11\, \times 10^{10}\, {\rm M_{\odot}}$ adopted from \cite{Licquia2015}.

Recent observations found that the derived perturbation time varies with the angular momentum $L_z$ or $R_g$ \citep[e.g.,][]{Antoja2023, Darragh-Ford2023, Frankel2023, Frankel2024, Widmark2026}. Therefore, we decided to use different $T_{evol}$ and $\theta_{z,0}$ for different radial bins of the observational data, i.e., 16 additional parameters for 8 radial bins. Including the 5 free parameters in the potential model, we have 21 free parameters in total. The priors adopted for the potential parameters are listed in Table \ref{tab:paras_obs}, while we adopt the same flat priors as listed in Table \ref{tab:paras_Irr} for all $T_{evol}$ and $\theta_{z,0}$. We notice that adopting different disk potential models, such as the exponential and double exponential disks, results in similar derived vertical potential profiles, local DM densities, perturbation times, and initial phase angles.

%%%%%%%%%%%%%%%%%%%%%%%%%%%%%%%%%%%%%%%%%
\begin{table*}
\centering
\caption{Parameters in the vertical potential measurement with the phase snails from {\gaia} DR3.}
\label{tab:paras_obs}
\begin{tabular}{cccccccl}
\hline
\hline
\multirow{2}{*}{\textbf{Parameter}} & \multirow{2}{*}{\textbf{Unit}} & \multirow{2}{*}{\textbf{Prior}} & \multicolumn{2}{c}{\textbf{Snails only}} & \multicolumn{2}{c}{\textbf{Snails + RC}} & \multirow{2}{*}{\textbf{Note}} \\
 &  &  & \textbf{Best-fit} & \textbf{Median} & \textbf{Best-fit} & \textbf{Median} &  \\
 (1) & (2) & (3) & (4) & (5) & (6) & (7) & (8) \\
\hline
\(M_b\) & $10^{10}{\rm M}_{\odot}$ & 1.067 & - & - & - & - & fixed \\
\(b_b\) & kpc & 0.3 & - & - & - & - & fixed \\
\(M_d\) & $10^{10}{\rm M}_{\odot}$ &  $5.17\pm1.11$ & 6.76 & $6.57^{+0.55}_{-0.65}$ & 6.34 & $7.18^{+1.15}_{-0.63}$ & fit \\
\(a_d\) & kpc &  [1, 8] & 6.10 & $4.37^{+0.68}_{-0.46}$ & 4.98 & $3.94^{+0.72}_{-0.29}$ & fit \\
\(b_d\) & kpc &  [0.05, 1] & 0.378 & $0.285^{+0.047}_{-0.041}$ & 0.315 & $0.306^{+0.014}_{-0.025}$ & fit \\
\(M_h\) & $10^{10}{\rm M}_{\odot}$ & [5, 300] & 53.9 & $51.3^{+11.1}_{-13.5}$ & 20.9 & $49.6^{+23.8}_{-26.1}$& fit \\
\(a_h\) & kpc & [1, 50] & 14.0 & $17.1^{+2.7}_{-2.8}$ & 5.8 & $14.3^{+4.7}_{-7.6}$ & fit \\
\hline
\(M_{200}\) & $10^{12}{\rm M}_{\odot}$ & - & 0.98 & $0.81^{+0.22}_{-0.23}$ & 0.51 & $0.87^{+0.31}_{-0.33}$ & derived \\
\(R_{200}\) & kpc & - & 205 & $193^{+16}_{-20}$ & 165 & $198^{+21}_{-29}$ & derived \\
\(c\) & - & - & 14.7 & $11.3^{+2.4}_{-1.9}$ & 28.5 & $13.9^{+11.2}_{-2.6}$ & derived \\
\(\rho_{dm,\odot}\) & ${\rm M_{\odot}\,pc^{-3}}$ & - & 0.0103 & $0.0076^{+0.0019}_{-0.0021}$ & 0.0100 & $0.0097^{+0.0002}_{-0.0009}$ & derived \\
\hline
\end{tabular}
\begin{tablenotes}
\small
\item \textbf{Notes.} Columns are: (1) Parameter names; (2) Units; (3) Priors adopt in the MCMC; (4) \& (5) Best-fit values and median values from all posterior parameter sets for the measurement utilizing only the snail shape curves; (6) \& (7) Best-fit values and median values from all posterior parameter sets for the measurement with the constraints from both the phase snails and the rotation curve; (8) Notes. Note that the scale length $a_d$ of a Miyamoto-Nagai disk is different from that ($R_d$) of an exponential disk, with the former slightly larger than the latter.
\end{tablenotes}
\end{table*}
%%%%%%%%%%%%%%%%%%%%%%%%%%%%%%%%%%%%%%%%%

%%%%%%%%%%%%%%%%%%%%%%%%%%%%%%%%%%%%%%%%%%%%%%%%%%%%%%%%%%%%%%
\section{Results} 
\label{sec:results}
%%%%%%%%%%%%%%%%%%%%%%%%%%%%%%%%%%%%%%%%%%%%%%%%%%%%%%%%%%%%%%
%%%%%%%%%%%%%%%%%%%%%%%%%%%%%%%%%%%%%%%%%
\subsection{Vertical potential measured from the snail shape curves only}
\label{ssec:obs_dTdtz}
\begin{figure*}[htbp]
\centering
\includegraphics[width=0.85\textwidth]{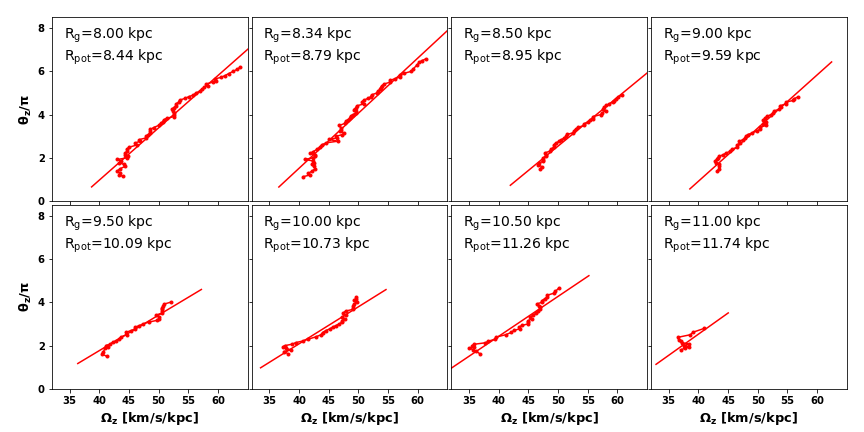}
\caption{\label{fig:Obs_Oztz_strong}$\Omega_{z} - \theta_{z}$ space for observed snail shape curves transformed under the best-fit potential from MCMC. The solid lines are the model linear relations based on the $T_{evol}$ and $\theta_{z,0}$ of each bin. The $R_{g}$ bin centers and the corresponding potential radii $R_{pot}$ are labeled on the upper left corner of each panel.}
\end{figure*}

%%%%%%%%%%%%%%%%%%%%%%%%%%%%%%%%%%%%%%%%%
We first measured the vertical potential with only the snail shape curves detected from {\gaia} DR3, utilizing the 1D orbital integration method and MCMC technique. The curves in $\Omega_{z} - \theta_{z}$ space converted from the $z-V_z$ phase snails under the best-fit potential from MCMC are shown in Fig. \ref{fig:Obs_Oztz_strong}. The curves are close to a linear relation at all radii, except for the outermost bin with fewer points and larger uncertainties. For the inner two bins, there are also slight deviations from the best-fit linear relations at both ends of the profiles, i.e., the smallest and largest $\Omega_z$. The points with small $\Omega_z$ correspond to the outermost region of $z-V_z$ space, where the numerical noise is large and the points are too close to the inner wrap. On the other hand, the points with large $\Omega_z$ locate close to the $z-V_z$ space central region, where the central clumpy feature influences the shape measurement slightly.

\begin{figure*}[htbp]
\centering
\includegraphics[width=0.85\textwidth]{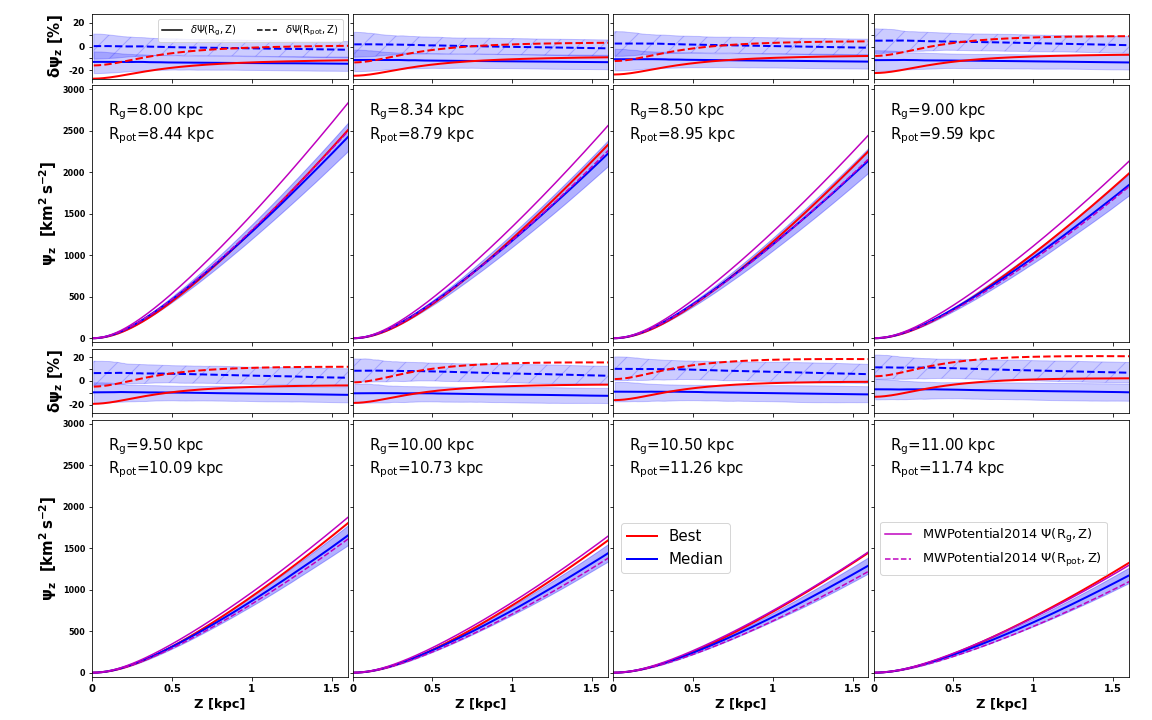}
\caption{
\label{fig:Obs_phiz_strong}
\textit{Second and fourth rows:} Vertical potential profiles measured from the observed snail shape curves. The red lines are the profiles from the best-fit potential of MCMC, while the blue lines are the median potential profiles from all posterior parameter sets, with the shaded regions indicating their 1$\sigma$ uncertainties. For comparisons, the vertical potentials from \cite{Bovy2015} (MWPotential2014) at the bin centers $R_g$ and the corresponding potential radii $R_{pot}$ are illustrated by the solid and dashed magenta lines, respectively. \textit{First and third rows:} Relative difference (as a percentage) between the vertical potential profiles measured from the observation and those at $R_g$ (solid) and $R_{pot}$ (dashed) of MWPotential2014.}
\end{figure*}

The model predicted vertical potential profiles are displayed in Fig. \ref{fig:Obs_phiz_strong}, compared to the popular Milky Way potential model MWPotential2014 from \cite{Bovy2015}. The best-fit potentials are indicated by the red lines, while the median profiles from all posterior parameter sets are shown by the blue lines with the shaded regions indicating the $1\sigma$ uncertainties. The vertical potentials at the bin centers $R_g$ and their corresponding potential radii $R_{pot}$ for MWPotential2014 are illustrated by the solid and dashed magenta lines, respectively. Our best-fit potential model shows good consistency with the potential profiles at $R_{pot}$ from MWPotential2014, especially at inner radii shown in the top row of Fig. \ref{fig:Obs_phiz_strong}. For the outer radii, the best-fit model prefers steeper potential profiles, and is more consistent with the potential at $R_g$ from MWPotential2014. This difference could result in the small curvatures of $\Omega_{z} - \theta_{z}$ profiles at the intermediate $\Omega_{z}$ for the outer three bins (at $\Omega_{z} \sim 45$ {\kms}\,$\rm kpc^{-1}$), shown in Fig. \ref{fig:Obs_Oztz_strong}. The linear relation and consistency in the potential profiles at inner radii mainly result from the better snail shape measurements with more data points and more wraps. Another possible reason is the large difference between $R_g$ and $R_{pot}$ at outer radii based on the empirical correction shown in Fig. \ref{fig:obs_Rpot}. The results seem to prefer a smaller potential radius correction there. Nevertheless, as shown in Figs. 9 and 10 of \cite{GuoR2024}, the potential profiles directly measured from the intersections between snails and $z/V_z$ axes also indicate a steeper vertical potential at these radii compared to the MWPotential2014 model. We have tried to select a colder observational sample utilizing a cut on the radial action, though it needs a prior potential to calculate actions. We found similar vertical potential measurements at inner radii, consistent with the model MWPotential2014 after the correction, while outer radii show slightly shallower potential profiles. We note that the outer region should be more carefully studied in the future, due to the larger noise, fewer wraps, and more susceptible to the influence of self-gravity.

%%%%%%%%%%%%%%%%%%%%%%%%%%%%%%%%%%%%%%%%%
%%%%%%%%%%%%%%%%%%%%%%%%%%%%%%%%%%%%%%%%%

%%%%%%%%%%%%%%%%%%%%%%%%%%%%%%%%%%%%%%%%%
%%%%%%%%%%%%%%%%%%%%%%%%%%%%%%%%%%%%%%%%%

\begin{figure}[bth]
\centering
\includegraphics[width=0.85\columnwidth]{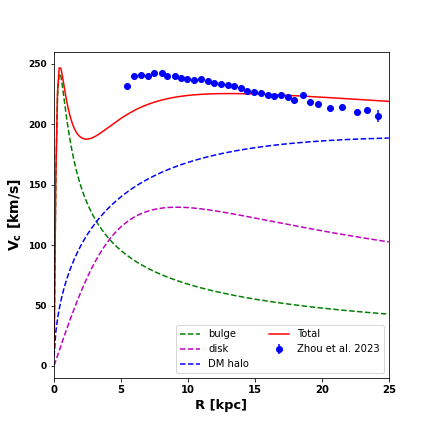}
\caption{
\label{fig:Obs_Vc_strong}
Rotation curve (red) from the best-fit potential. The dashed lines are the contributions from the different components: green for the bulge, magenta for the disk, and blue for the DM halo. The blue dots are the shifted rotation curve measurements from \cite{ZhouY2023}.}
\end{figure}

Although our best-fit model shows a good linear performance in $\Omega_{z} - \theta_{z}$ space and a good consistency in the measured vertical potential compared to the MWPotential2014, the rotation curve from the best-fit model is not consistent with that from \cite{ZhouY2023}, as shown in Fig. \ref{fig:Obs_Vc_strong}. This recent rotation curve measurement utilizing the radial Jeans equation shows a weak decline as $R$ with a gradient of $-1.83$ {\kms}\,kpc$^{-1}$ up to $R=25$ kpc, while our rotation curve is flatter. This difference results from the fact that the radial range explored here is only $8 \le R \le 11$ kpc, leading to a weak constraint on the radial variation in the rotation curve. It also leads a loose constraint on the DM halo properties. The best-fit potential model indicates a DM halo with $M_{200}= 0.98\times10^{12}\, {\rm M_{\odot}}$, $R_{200}= 205$ kpc, a concentration parameter of $c= 14.7$ and a local DM density of $\rho_{dm,\odot}= 0.0103\,{\rm M_{\odot}\,pc^{-3}}$. The median values and $1\sigma$ uncertainties of these derived parameters from all MCMC posterior parameters sets are: $M_{200}= 0.81^{+0.22}_{-0.23} \times10^{12}\, {\rm M_{\odot}}$, $R_{200}= 193^{+16}_{-20}$ kpc, $c= 11.3^{+2.4}_{-1.9}$, $\rho_{dm,\odot}= 0.0076^{+0.0019}_{-0.0021} \,{\rm M_{\odot}\,pc^{-3}}$. The derived DM mass of Milky Way is consistent with recent measurements from rotation curve measurements: $0.805\pm0.115\times10^{12}\, {\rm M_{\odot}}$ from \cite{ZhouY2023}, $0.725\pm0.026\times10^{12}\, {\rm M_{\odot}}$ from \cite{Eilers2019}, $0.725\pm0.025\times10^{12}\, {\rm M_{\odot}}$ from \cite{WangHF2023}, and $0.692^{+0.012}_{-0.011} \times10^{12}\, {\rm M_{\odot}}$ from \cite{OuXW2024}. Note that our best-fit value of $M_{200}$ is higher than the median value of its marginalized distribution. Its flat rotation curve also indicates a larger $M_{200}$ than the reference works with declining rotation curves. In addition, we also note that the adoption of different DM models can result in significantly different $M_{200}$ estimations \citep{JiaoYJ2021, OuXW2024, SunGC2025}, especially for the Einasto profile versus the NFW DM profile.

%%%%%%%%%%%%%%%%%%%%%%%%%%%%%%%%%%%%%%%%%
%%%%%%%%%%%%%%%%%%%%%%%%%%%%%%%%%%%%%%%%%

%%%%%%%%%%%%%%%%%%%%%%%%%%%%%%%%%%%%%%%%%
\subsection{Combined model fitting by including the rotation curve}
\label{ssec:obs_with_Vc}
\begin{figure*}[htbp]
\centering
\includegraphics[width=0.85\textwidth]{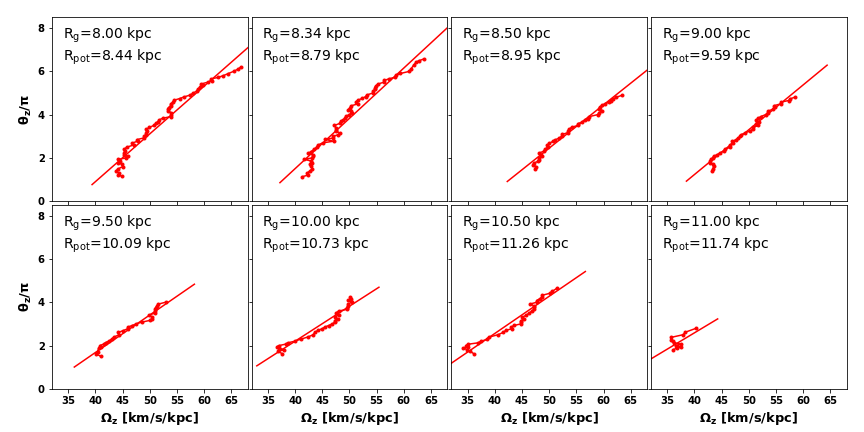}
\caption{
\label{fig:Obs_Oztz_Vc}Same as Fig. \ref{fig:Obs_Oztz_strong} but for the modeling with the constraints from both the phase snails and the rotation curve.}
\end{figure*}

\begin{figure*}[htbp]
\centering
\includegraphics[width=0.85\textwidth]{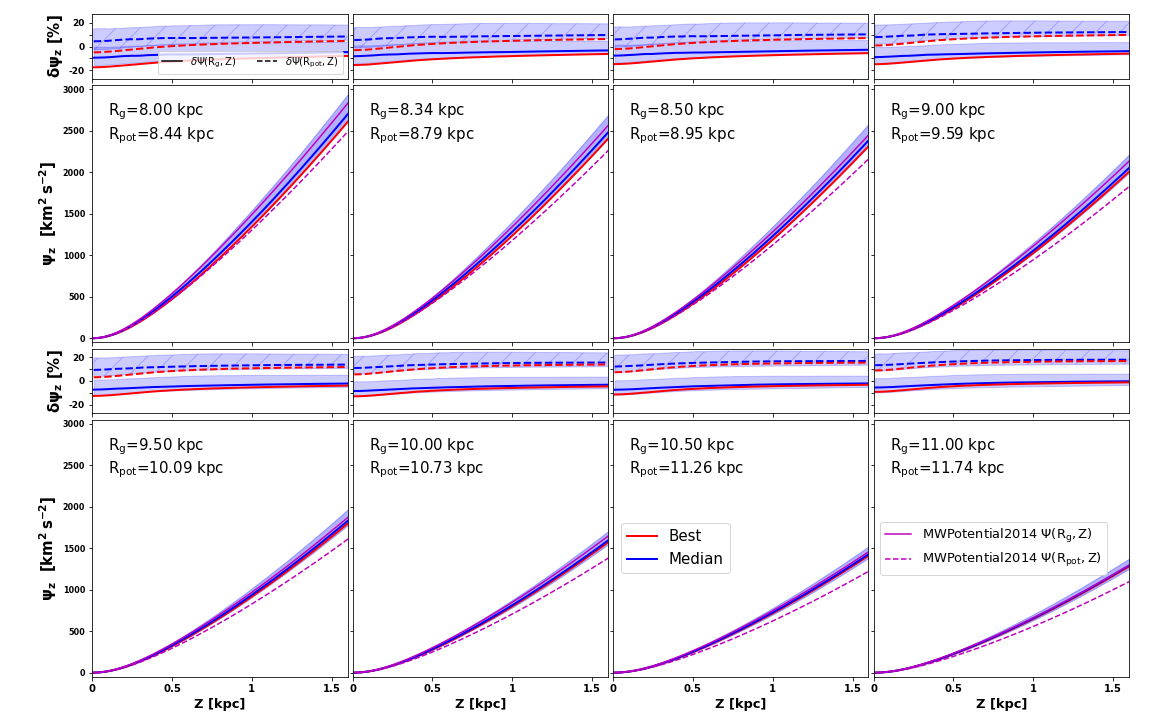}
\caption{
\label{fig:Obs_phiz_Vc}Same as Fig. \ref{fig:Obs_phiz_strong} but for the modeling with the constraints from both the phase snails and the rotation curve.}
\end{figure*}

\begin{figure}[bth]
\centering
\includegraphics[width=0.85\columnwidth]{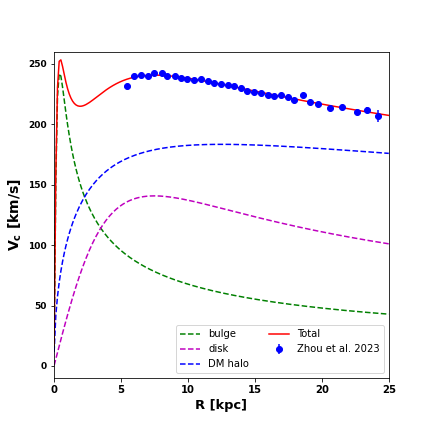}
\caption{
\label{fig:Obs_Vc_Vc}Same as Fig. \ref{fig:Obs_Vc_strong} but for the modeling with the constraints from both the phase snails and the rotation curve.}
\end{figure}

%%%%%%%%%%%%%%%%%%%%%%%%%%%%%%%%%%%%%%%%%
Here we combine both the observed Galactic rotation curve with the vertical phase snails to constrain the Milky Way potential model. The rotation curve provides a stronger constraint on the mass distribution in the radial direction, while the phase snails offer stronger constraints in the vertical direction. As already mentioned before, we modified the rotation curve in \cite{ZhouY2023} to compensate for the difference in the solar radius $R_0$. The converted $\Omega_{z} - \theta_{z}$ curves under the best-fit model, the model predicted vertical potential profiles and the rotation curve from the best-fit model are shown in Figs. \ref{fig:Obs_Oztz_Vc}, \ref{fig:Obs_phiz_Vc}, and \ref{fig:Obs_Vc_Vc}, respectively. The best-fit model with constraints from both the phase snails and the rotation curve results in roughly similar linearity of the curves in $\Omega_{z} - \theta_{z}$ space. The model predicts steeper vertical potential profiles than those derived from the snail shape curves only, as indicated by the shaded regions in Fig. \ref{fig:Obs_phiz_Vc}. The best-fit model is now more consistent with the median profiles from all posterior parameter sets, which are more consistent with the vertical potential at $R_{g}$ from the MWPotential2014 model, rather than those at $R_{pot}$. This discrepancy could be due to the weakly declining observed rotation curve, which leads to larger inner total densities, and thus steeper inner vertical potential profiles, and also a faster radial variation (smaller disk radial length).

%%%%%%%%%%%%%%%%%%%%%%%%%%%%%%%%%%%%%%%%%
%%%%%%%%%%%%%%%%%%%%%%%%%%%%%%%%%%%%%%%%%

%%%%%%%%%%%%%%%%%%%%%%%%%%%%%%%%%%%%%%%%%
%%%%%%%%%%%%%%%%%%%%%%%%%%%%%%%%%%%%%%%%%

%%%%%%%%%%%%%%%%%%%%%%%%%%%%%%%%%%%%%%%%%
%%%%%%%%%%%%%%%%%%%%%%%%%%%%%%%%%%%%%%%%%

With the additional constraint from observed rotation curve, the rotation curve from our best-fit model is now consistent with the observation, as shown in Fig. \ref{fig:Obs_Vc_Vc}. In the best-fit mass model, the mass of the DM halo of Milky Way is $M_{200}= 0.51\times10^{12}\, {\rm M_{\odot}}$, and the corresponding radius is $R_{200}= 165$ kpc, with a concentration parameter of $c= 28.5$ and a local DM density of $\rho_{dm,\odot}= 0.0100\,{\rm M_{\odot}\,pc^{-3}}$. The median values and $1\sigma$ uncertainties of these derived parameters from all MCMC posterior parameters sets are: $M_{200}= 0.87^{+0.31}_{-0.33} \times10^{12}\, {\rm M_{\odot}}$, $R_{200}= 198^{+21}_{-29}$ kpc, $c= 13.9^{+11.2}_{-2.6}$, $\rho_{dm,\odot}= 0.0097^{+0.0002}_{-0.0009} \,{\rm M_{\odot}\,pc^{-3}}$. The DM halo derived by combining both the rotation curve and phase snails has similar $M_{200}$ and $R_{200}$, and a slightly larger concentration $c$. Their uncertainties are also larger than those derived utilizing only the snails, in contrast to a significantly decreased uncertainty of $\rho_{dm,\odot}$. The larger uncertainties in $M_{200}$ and $R_{200}$ are due to the double peaks in the posterior distributions of $M_h$ and $a_h$ in this combined model, with one peak at $M_h \sim 25\times10^{10}\, {\rm M_{\odot}}$ and $a_h \sim 7.5$ kpc, while another at $M_h \sim 67\times10^{10}\, {\rm M_{\odot}}$ and $a_h \sim 18.1$ kpc. Nevertheless, these two parts of posterior parameters have precise and slightly discrepant values of $\rho_{dm,\odot}$, with median values and $1\sigma$ uncertainties of $\rho_{dm,\odot}= 0.0099^{+0.0001}_{-0.0002} \,{\rm M_{\odot}\,pc^{-3}}$ and $\rho_{dm,\odot}= 0.0091^{+0.0007}_{-0.0004} \,{\rm M_{\odot}\,pc^{-3}}$. The convergence in $\rho_{dm,\odot}$ for different parameter sets results in its high precision, which could be due to the strong constraint resulting from the combination of the rotation curve and snails near solar radii. Large discrepancy emerges at outer radii ($R\gtrsim 15$ kpc), as the mass model derived using only the snails ($8 \le R_g \le 11$ kpc) prefers a flatter rotation curve than that of the observation, as shown in Fig. \ref{fig:Obs_Vc_strong}, leading to the double peaks in posterior distributions.

The increased constraint on $\rho_{dm,\odot}$ under the combination of rotation curve is independent of the disk model adopted. Adopting an exponential disk, the combined model results in a more precise value of $\rho_{dm,\odot}= 0.0094^{+0.0006}_{-0.0004} \,{\rm M_{\odot}\,pc^{-3}}$, compared to $\rho_{dm,\odot}= 0.0104^{+0.0023}_{-0.0030} \,{\rm M_{\odot}\,pc^{-3}}$ for the model using snails only. For the double exponential disks model, these values are $\rho_{dm,\odot}= 0.0090^{+0.0005}_{-0.0003} \,{\rm M_{\odot}\,pc^{-3}}$ and $\rho_{dm,\odot}= 0.0108^{+0.0023}_{-0.0021} \,{\rm M_{\odot}\,pc^{-3}}$, respectively\footnote{For the exponential disk model, we adopt the same Gaussian prior of $M_d=5.17\pm1.11\, \times 10^{10}\, {\rm M_{\odot}}$ \citep{Pouliasis2017}, while for the double exponential disks model, we adopt priors of total disk mass of $M_{d,thin} + M_{d,thick}= 5.17\pm1.11\, \times 10^{10}\, {\rm M_{\odot}}$, surface density ratio at solar radius of $\Sigma_{thick,\odot}/\Sigma_{thick,\odot}= 12\pm4\%$ \citep{Bland-Hawthorn2016}, and fixed disk scale heights of $z_{d,thin}= 0.3$ kpc and $z_{d,thick}= 0.9$ kpc.}. Including the rotation curve increases the uncertainty of $\rho_{dm,\odot}$ from $\sim 0.0025$ to $\sim 0.0005 \,{\rm M_{\odot}\,pc^{-3}}$. Besides, fitting only the rotation curve results in $\rho_{dm,\odot}  \sim 0.01 \,{\rm M_{\odot}\,pc^{-3}}$, but nearly with no constraints on the disk scale heights and the vertical potential. Note that we do not aim to constrain the parameters of the thick disk with the current data, these tests just attempt to demonstrate that the snails can constrain the total vertical potential profiles and thus result in similar $\rho_{dm,\odot}$ measurements. The different disk models also result in similar perturbation time measurements shown in the next section.

%%%%%%%%%%%%%%%%%%%%%%%%%%%%%%%%%%%%%%%%%
\subsection{Constraint on the perturbation time}
\label{ssec:obs_Tevol}
\begin{figure}[bth]
\centering
\includegraphics[width=0.9\columnwidth]{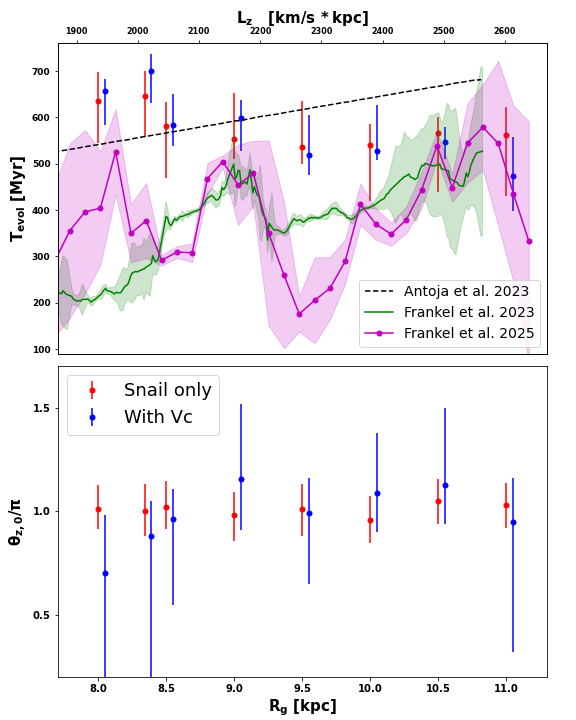}
\caption{
\label{fig:Obs_diff_Tevol}Perturbation time, $T_{evol}$ (\textit{top}), and initial phase angle, $\theta_{z,0}$ (\textit{bottom}), as functions of $R_g$ for the modeling utilizing the snail shape curves only (red dots) or combining the constraints from the phase snails and rotation curve (blue dots). The blue dots are slightly shifted for clarity. The corresponding $L_z$ is indicated in the top $X$ axis. The dashed black line displays the global increasing trend from the linear fit to the measurements of \cite{Antoja2023}, while the green and magenta lines show the measurements from \cite{Frankel2023} and \cite{Frankel2024}, respectively, utilizing different catalogs and dynamical analyses of the Galactic potential. All these measurements are shifted according to our local angular momentum, $L_{z,\odot}= R_{0}\times V_{c,\odot}$.}
\end{figure}

%%%%%%%%%%%%%%%%%%%%%%%%%%%%%%%%%%%%%%%%%
Except for the potential parameters, our method also provides constraints on the perturbation evolution time, $T_{evol}$, and the initial phase angle, $\theta_{z,0}$. The estimations of $T_{evol}$ and $\theta_{z,0}$ at different $R_g$ are displayed in Fig. \ref{fig:Obs_diff_Tevol}. The red dots are the measurements of the modeling utilizing the snail shape curves only, while the blue dots are the predictions from the combined constraints of the phase snails and rotation curve. The two models are consistent with each other at most radii, though with different uncertainties.

%%%%%%%%%%%%%%%%%%%%%%%%%%%%%%%%%%%%%%%%%
%%%%%%%%%%%%%%%%%%%%%%%%%%%%%%%%%%%%%%%%%

The overall measurement of $T_{evol}$ for both models is in the range of 400-750 Myr with a median of 564 Myr, which is consistent with several recent works utilizing different methods \citep{Widmark2022a, Antoja2023, Darragh-Ford2023, Frankel2023, Frankel2024}, though showing a slightly decreasing pattern with $R_g$ or $L_z$. The different methods and different potential models adopted could result in the different variations with $L_z$ or $R_g$. We compare our measurements with recent works and discuss the differences in Sect. \ref{ssec:dis_pert}. The median values of $\theta_{z,0}$ are always close to $\pi$, perhaps indicating a perturbation similar to a downward vertical velocity kick. The uncertainties of $\theta_{z,0}$ are larger when combining the constraints from both the phase snails and the rotation curve, compared to those in the model using snails only. It has a similar origin to the increased uncertainties in $M_h$ and $a_h$, i.e., the double peaks in the posterior distributions, as $T_{evol}$ and $\theta_{z,0}$ are strongly degenerate with the potential parameters.

%%%%%%%%%%%%%%%%%%%%%%%%%%%%%%%%%%%%%%%%%%%%%%%%%%%%%%%%%%%%%%
\section{Discussion}
\label{sec:discuss}
%%%%%%%%%%%%%%%%%%%%%%%%%%%%%%%%%%%%%%%%%%%%%%%%%%%%%%%%%%%%%%
%%%%%%%%%%%%%%%%%%%%%%%%%%%%%%%%%%%%%%%%%
\subsection{Method comparison with previous works}
\label{ssec:dis_comp}
%%%%%%%%%%%%%%%%%%%%%%%%%%%%%%%%%%%%%%%%%
This work aims to measure the vertical potential with the snail shape curves. It is a direct extension of our previous work \citep{GuoR2024}, in which we utilized the intersections between phase snails and $z/V_z$ axes and a refined linear interpolation method to directly measure the vertical potential at these intersections. Although the method in \cite{GuoR2024} is model-independent, it leverages very little information of the snails, which naturally encode the information about the vertical potential in their shapes. Therefore, in this work, we utilize the snail shape curves and the 1D phase-space orbital integration inspired by \cite{Widmark2021a} to measure the vertical potential, focusing on the linear pattern of the phase snail in the $\Omega_{z} - \theta_{z}$ space.

Our work combines the methods used for snail modeling, i.e., 1D phase-space orbital integration \citep{Widmark2021a, Widmark2021b}, and snail shape extraction \citep{LiZY2021, Darragh-Ford2023}. There are two main differences between our method and \cite{Widmark2021a}. First, we focused on the linear relationship of the $\Omega_{z} - \theta_{z}$ curve instead of constructing the analytical function of phase snail. Second, we extracted the snail shape curve from the 2D phase-space density map first and then measured the vertical potential utilizing the extracted shape curves, while \cite{Widmark2021a} directly fit the 2D phase-space distribution of the residual density map. Their 2D fitting is time-consuming due to the numerical integration for all stars, and is more sensitive to the noise from the bulk background and substructures (e.g., secondary branches). In \cite{Darragh-Ford2023}, the snail shape was extracted in $\theta_z - \sqrt{J_z}$ space after they fixed the potential model, which was based on a visual inspection of the linear performance of one local snail in $\theta_z - \sqrt{J_z}$ space (see their Appendix A for more details). They only examined $5\times5$ parameter grids for the disk scale height and disk mass. This is because action calculations are highly time-consuming when performed for all observed stars, and computational costs would escalate further if more parameter combinations were included. In contrast, our adopted strategy circumvents this limitation, while ensuring accuracy: we first extracted the snail-shaped curves from the $z-V_z$ phase space (each snail is typically characterized by a large number of data points), then conducted 1D phase-space orbital integrations to constrain the underlying gravitational potential. This approach efficiently balances computational feasibility with precision, thereby enabling us to employ the MCMC technique for an exhaustive exploration of a much broader parameter space.

Several other works made use of the vertical phase-space kinematics (or with additional element abundances)  to estimate the vertical potential under equilibrium state, with the snails showing in the residuals between the model and observation. \cite{LiHC2021, LiHC2023} utilized the rational linear distribution function model to directly fit the number density in $z-V_z$ plane, and thus determined the vertical potential and force. Snails emerge in the residual maps of $z-V_z$ plane, which lead to an estimation of the perturbation time of $\sim 540$ Myr. Orbital Torus Imaging, a new dynamical inference technique, utilizes kinematic measurements and element abundances (or other invariant labels) to infer the Galactic potential \citep{Price-Whelan2021, Horta2024, Price-Whelan2025}. The basic idea is that stellar labels vary systematically with orbit actions, while invariant with respect to orbital phases (conjugate angles). The best-fit mass model is the one leading to no residual dependence of abundances on phase angles. Faint snail features also show in the abundance residuals between the model and data, or clearly present in the maps after subtracting the trend (or gradient) between the mean metallicity (stellar age or mass) and orbit action, which are called Iron Snails \citep{Frankel2024}. The snails could bias the inferred orbit shapes, and thus the acceleration profiles, which could be solved by explicitly modeling the disk perturbation and its dependence on stellar kinematics and abundances in the future.

%%%%%%%%%%%%%%%%%%%%%%%%%%%%%%%%%%%%%%%%%
\subsection{Comparisons of vertical potential and vertical force}
\label{ssec:dis_paras}
\begin{figure}[bth]
\centering
\includegraphics[width=0.9\columnwidth]{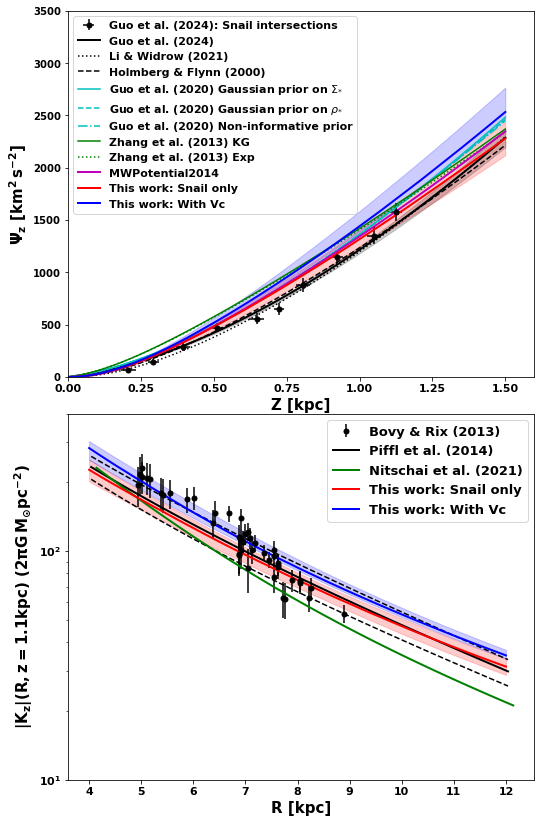}
\caption{
\label{fig:comp_phiz_Sigma}Comparisons of the vertical potential profile at solar radius (\textit{top}) and the vertical force at 1.1 kpc as a function of radius (\textit{bottom}), i.e., $|K_z|(R,z=1.1\ {\rm kpc})$. The red and blue lines are our predictions from models using snails only and by combining the snails and rotation curve, respectively, with the shaded regions indicating uncertainties.}
\end{figure}

%%%%%%%%%%%%%%%%%%%%%%%%%%%%%%%%%%%%%%%%%
The vertical potential is well recovered through our method for the test particle simulation, as shown in Fig. \ref{fig:simu_phiz}. Moreover, the potential profiles measured from the observational data are also well consistent with the model MWPotential2014 \citep{Bovy2015} as shown by Figs. \ref{fig:Obs_phiz_strong} and \ref{fig:Obs_phiz_Vc}. We also compare the vertical potential profiles at solar radius measured from several local dynamic studies with those derived from our method, displaying in the upper panel of Fig. \ref{fig:comp_phiz_Sigma}. The vertical potential derived using snails only, i.e., the red line with the shaded region, is consistent with the profiles from \cite{Holmberg2000}, \cite{LiHC2021}, and also the measurements from our previous work utilizing the snail intersections \citep{GuoR2024}, i.e., the black lines and points. The profile derived under the combined constraints from the snails and rotation curve, i.e., the blue line with the shaded region, is systematically steeper than the red line. It is also more consistent with the profiles derived in \cite{ZhangL2013}. The profiles derived utilizing vertical Jeans equation in \cite{GuoR2020}, i.e., the cyan lines, lie between the two profiles from this work.

The lower panel of Fig. \ref{fig:comp_phiz_Sigma} shows the comparison of the vertical force at $z=1.1$ kpc as a function of radius, i.e., $|K_z|(R,z=1.1\ {\rm kpc})$ from our work and other three works. This vertical force is the first order derivative of the vertical potential at 1.1 kpc, and is similar to the total surface density within 1.1 kpc of the mid-plane (equal to that under a flat rotation curve). The red and blue lines (with shaded regions indicating $1\sigma$ uncertainties) show the model predicted vertical force profiles derived using snails only and by combining snails and the rotation curve, respectively. The black points are the measurements by fitting a parameterized Milky Way potential and an action-based distribution function to the phase-space data of different mono-abundance populations \citep{Bovy2013}. The solid black line (dashed lines indicating its uncertainties) shows the result for the dynamical fit to 200,000 giants in the RAVE survey \citep{Piffl2014}, while the green line displays the model prediction from a data set that combines Gaia EDR3 and APOGEE utilizing the spherically aligned Jeans anisotropic method \citep{Nitschai2021}. The $K_z$ profile derived using snails only is well consistent with the profile from \cite{Piffl2014}, consistent with measurements from \cite{Bovy2013} in the region of $R> 7$ kpc, and slightly larger than that of \cite{Nitschai2021} at $R>8$ kpc. Interestingly, the systematically larger $K_z$ profile from our combined model with the additional constraint from rotation curve is consistent with \cite{Bovy2013} at $R<7$ kpc. Note that our measurements are based on the data at $8<R<11$ kpc. The consistency of $|K_z|(R,z=1.1\ {\rm kpc})$ between our measurements and other works provides an independent validation of our work.

%%%%%%%%%%%%%%%%%%%%%%%%%%%%%%%%%%%%%%%%%
%%%%%%%%%%%%%%%%%%%%%%%%%%%%%%%%%%%%%%%%%

%%%%%%%%%%%%%%%%%%%%%%%%%%%%%%%%%%%%%%%%%
\subsection{Comparisons of DM halo quantities}
\label{ssec:dis_paras}
\begin{figure*}[htbp]
\centering
\includegraphics[width=0.98\textwidth]{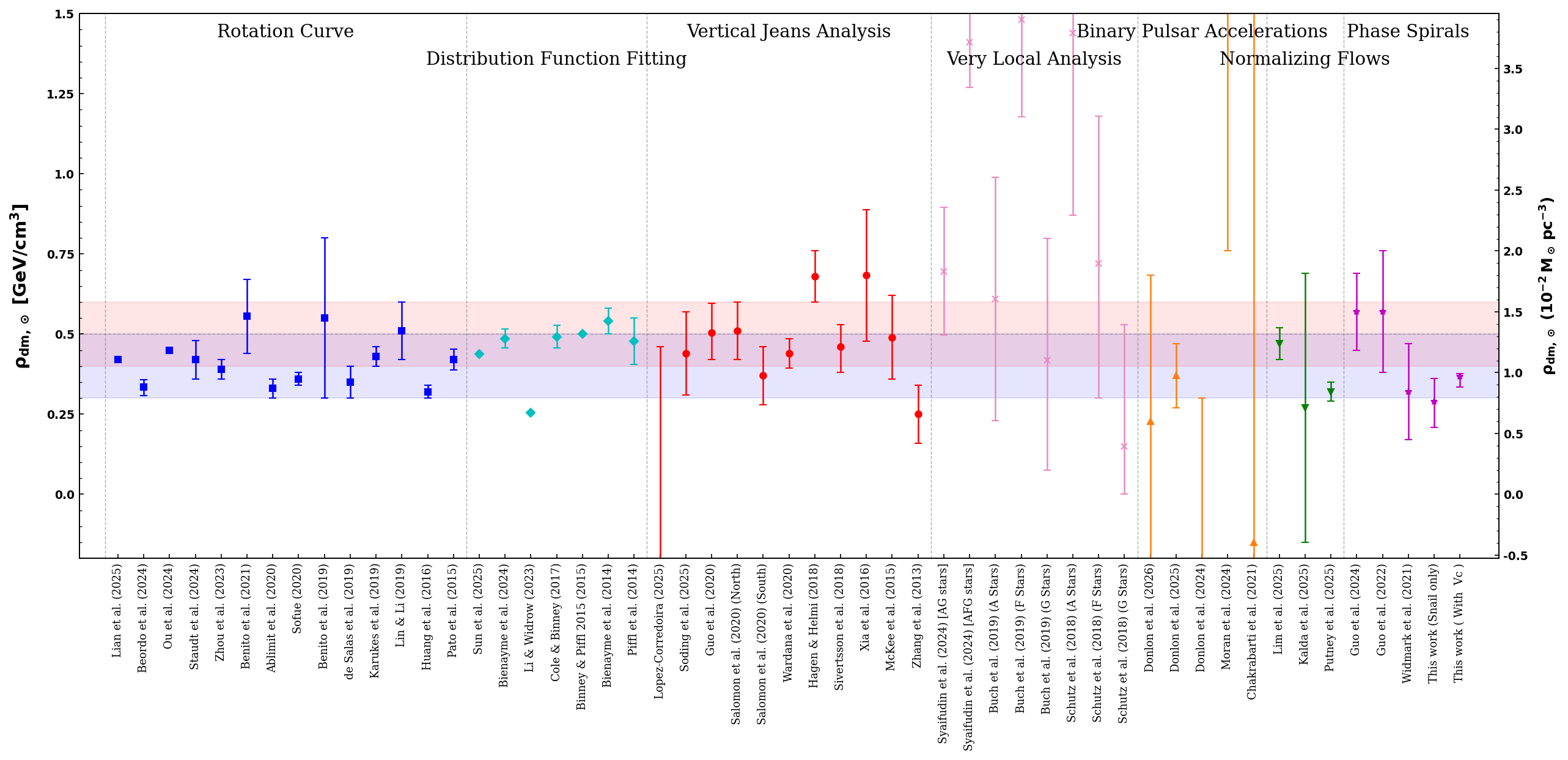}
\caption{
\label{fig:comp_rhodm}Comparison of the local DM density, $\rho_{dm,\odot}$, with recent literature. The measurements are divided into seven categories: blue squares are measurements based on global fitting of the rotation curve; cyan diamonds are measurements based on distribution function fitting; red circles are measurements using vertical Jeans analysis; pink crosses are works based on very local analysis; orange triangles are works based on binary pulsar accelerations; green triangles are measurements using normalizing flows; and magenta stars are measurements using the vertical phase spirals. The rightmost two magenta stars indicate the measurements in this work. The shaded blue and red regions show the ranges of measurements based on global studies and local analyses, respectively, according to the compilation in \cite{deSalas2021}.}
\end{figure*}

%%%%%%%%%%%%%%%%%%%%%%%%%%%%%%%%%%%%%%%%%
Local DM density, as a basic parameter of the Milky Way mass model, provides an important input for DM direct detection experiments. Traditional measurements usually can be separated into several categories. The first popular category is the global mass model fitting to the observed rotation curve \citep[e.g.,][]{Pato2015, HuangY2016, deSalas2019, LinHN2019, Karukes2019, Benito2019, Sofue2020, Ablimit2020, Benito2021, ZhouY2023, OuXW2024, Staudt2024, Beordo2024, LianJH2025}. The second popular category is usually based on the local stellar kinematics using the vertical Jeans analysis \citep[e.g.,][]{ZhangL2013, McKee2015, XiaQR2016, Hagen2018, Sivertsson2018, Wardana2020, GuoR2020, Salomon2020, Lopez-Corredoira2025, Soding2025}. The third category analyzes the disk stars applying a distribution function fitting approach, which provides global Galactic mass models compatible with observations \citep[e.g.,][]{Piffl2014, Bienayme2014, Binney2015, Cole2017, LiHC2023, Bienayme2024, SunGC2025}. The fourth category applies 1D distribution function method to fit observed velocity or number density distributions based on very local analyses (usually $|z|< 200-300$ pc) of different types of stars, which usually have large uncertainties and discrepancies \citep[e.g.,][]{Schutz2018, Buch2019, Syaifudin2024}. The large uncertainty for these very local analyses could be due to the complicated influence of phase spirals to the very inner region of phase space. Since 2020, various new methods have been proposed based on the analysis of the binary pulsar accelerations \citep[e.g.,][]{Chakrabarti2021, Moran2024, Donlon2024, Donlon2025, Donlon2026}, the machine learning technique using normalizing flows \citep[e.g.,][]{Kalda2025, Lim2025, Putney2025}, or the newly found phase spirals \citep[e.g.,][]{Widmark2021b, GuoR2022, GuoR2024}. The local DM density is an important derived quantity of this work. We provide a comparison of the local DM density with literature in Fig. \ref{fig:comp_rhodm}. For more detailed comparisons of the local DM density measurements from different methods, we refer the readers to the review article \cite{deSalas2021}, and a recent compilation (Fig. 7) by \cite{Putney2025}.

Our measurements are $\rho_{dm,\odot}= 0.0076^{+0.0019}_{-0.0021} \,{\rm M_{\odot}\,pc^{-3}}$ from snails only and $\rho_{dm,\odot}= 0.0097^{+0.0002}_{-0.0009} \,{\rm M_{\odot}\,pc^{-3}}$ by combining snails and the rotation curve, consistent with many previous works. We state that the combination of the rotation curve and the phase snails should provide strong two directional constraints on the Milky Way potential. We do obtain a local DM density with very small uncertainties under only a loose Gaussian prior of disk mass $M_d$ when combining constraints from the rotation curve and phase snails, compared to the larger uncertainty of $\sim 0.002\, {\rm M_{\odot}\,pc^{-3}}$ for the measurement using snails only.

We tried different methods to measure the local DM density. \cite{GuoR2020} used the traditional vertical Jeans equation and the local kinematics to measure the local DM density. \cite{GuoR2022} applied a superposition of a smooth background and a phenomenal snail model to explain the north-south asymmetry found in \cite{GuoR2020}, while the local DM density is measured based on the vertical Jeans modeling of the smooth background under equilibrium assumption. Furthermore, in \cite{GuoR2024}, we obtained a local DM density measurement through the mass modeling of the direct potential measurements from the snail intersections at the solar radius. These works result in $\rho_{dm,\odot}$ measurements in a range of $0.013-0.015 \,{\rm M_{\odot}\,pc^{-3}}$, which are higher than those in this work.

The difference could be due to the fact that the current work is based on the global analyses of snails at different radii, while the previous three works rely solely on the vertical kinematics of the solar neighborhood. This difference between the global and local analyses is similar to the discrepancy found in the review article \cite{deSalas2021}, shown in their Fig. 1. They compared studies on measuring $\rho_{dm,\odot}$, and found that the results of local analyses coincide with a range of $\rho_{dm,\odot}\simeq 0.011-0.016 \,{\rm M_{\odot}\,pc^{-3}}$, while most global studies prefer a slightly lower range of $\rho_{dm,\odot}\simeq 0.008-0.013 \,{\rm M_{\odot}\,pc^{-3}}$, corresponding to the shaded red and blue regions in Fig. \ref{fig:comp_rhodm}, respectively. In addition, in our previous three works, we assumed a flat rotation curve in the simplified vertical Jeans equation. A declining rotation curve will contribute a negative correction of $\frac{1}{4\pi GR} \frac{\partial V_c^2(R)}{\partial R}$ to the measurements. Considering the recent negative radial gradients of the observed rotation curves \citep{Eilers2019, JiaoYJ2021, WangHF2023, ZhouY2023, OuXW2024}, $\rho_{dm,\odot}$ from ours and some other (not all) local kinematic studies should be $0.002-0.003 \,{\rm M_{\odot}\,pc^{-3}}$ lower. This correction makes our previous local studies closer to the current work, partially alleviating the difference between local and global analyses. Nevertheless, the residual difference may provide an indication of the DM halo shape.

Our model predicted halo mass, $M_{200}$, is also consistent with recent works based on Milky Way rotation curves \citep[e.g.,][]{Eilers2019, WangHF2023, ZhouY2023, OuXW2024}. The measurement using snails only ($M_{200}= 0.81^{+0.22}_{-0.23} \times10^{12}\, {\rm M_{\odot}}$) should place a loose constraint on the halo mass due to the small data coverage, as also illustrated by the loose constraints on $M_h,\ a_h$ in the test particle simulation. Nevertheless, this measurement is consistent with that from the model with additional constraint from the observed rotation curve ($M_{200}= 0.87^{+0.31}_{-0.33} \times10^{12}\, {\rm M_{\odot}}$), as both measurements have large uncertainties. Although, the derived local DM density of the latter model has smaller uncertainties, the uncertainties of other DM parameters ($M_{200}, R_{200}, c$) are larger than those of the former model. The increased uncertainties, manifested as the double peaks in the posterior distributions, could originate from the discrepancy at large radii ($R\gtrsim 15$ kpc). Specifically, the mass model derived from the snails only ($8 \le R_g \le 11$ kpc) prefers a flatter rotation curve compared to the observation. In addition, the empirical correction of potential radius $R_{pot}$ for $R_g-$binned snails, which would influence the radial variation in the vertical potential, could also contribute to this discrepancy. According to the compilation (Fig. 1) in the review article \cite{WangWT2020}, most measurements of $M_{200}$ (including methods based on Galactic escape velocity, rotation curve, spherical Jeans equation, distribution function, and stellar streams) are in the range of $0.5-2 \times10^{12}\, {\rm M_{\odot}}$. Our measurements are on the lower side of this range. See more detailed comparisons with other works in \cite{WangWT2020}.

%%%%%%%%%%%%%%%%%%%%%%%%%%%%%%%%%%%%%%%%%
%%%%%%%%%%%%%%%%%%%%%%%%%%%%%%%%%%%%%%%%%

%%%%%%%%%%%%%%%%%%%%%%%%%%%%%%%%%%%%%%%%%
\subsection{Comparing the perturbation time with previous works}
\label{ssec:dis_pert}
%%%%%%%%%%%%%%%%%%%%%%%%%%%%%%%%%%%%%%%%%
Our estimation of $T_{evol}$ ranges from 400 to 750 Myr with a median of 564 Myr, showing a slightly decreasing trend with $R_g$/$L_z$. We compare several recent measurements with ours in Fig. \ref{fig:Obs_diff_Tevol}, with their angular momenta $L_z$ shifted according to our local angular momentum $L_{z,\odot}= R_{0}\times V_{c,\odot}$. Our method is similar to that used in \cite{Widmark2022a}, which also focus on the potential measurement and provided an estimation of $T_{evol}$ of $622^{+145}_{-272}$ Myr (at $R\sim 9.8$ kpc) in their appendix. \cite{Antoja2023} utilize the vertical frequency difference of consecutive turn-over points of snails to estimate the perturbation time (also in \citealt{Antoja2018}). Their measurement is in a range of 300-900 Myr, although different potential models could result in a systematic difference of about 200 Myr. They find robust tendencies of $T_{evol}$ increasing with $L_z$ and $z$. \cite{Darragh-Ford2023} extract the snails in $J_z - \theta_z$ space and fit the extracted points in  $\Omega_z - \theta_z$ space with a line. Their estimated $T_{evol}$ range from 288 to 966 Myr with a median of 524 Myr, showing an obvious peak at $L_z\sim2150$ kpc\,{\kms} ($R_g \sim 9.3$ kpc). The measurements of \cite{Frankel2023} and \cite{Frankel2024} are in a range of 200-600 Myr, showing similar peaks and variation with $L_z$. These two works utilized completely different dynamical analyses: the former relies on the implementation of Milky Way potential (MWPotential2014), while the latter relies on empirical measurements of orbit shapes of Iron Snails with the Orbital Torus Imaging method \citep{Price-Whelan2021}. 

The estimation of $T_{evol}$ significantly depends on the Galactic potential model, the calculation of the vertical frequency (1D vs. 3D), the snail shape modeling, and the guiding radius calculation. Compared to the previous works, the different variation trends with $L_z$, and the difference of $T_{evol}$ at $R_g <8.5$ kpc (or $L_z< 2000$ kpc\,{\kms}) could be due to these complicated factors. Nevertheless, a peak at $L_z\sim2150$ kpc\,{\kms} seems to emerge in various studies \citep{Antoja2023, Frankel2023, Frankel2024, Darragh-Ford2023}. In addition, the self-gravity and other damping effect due to the sources such as giant molecular clouds may bias the measured perturbation time to lower values \citep{Widrow2023, Asano2026}. \cite{Asano2026} found the self-gravitating Galactic disk may delay the phase mixing by $\sim 0.3$ Myr in the solar neighborhood, under an interaction with the Sagittarius. Accounting for this delay revises the inferred $T_{evol}$ to $\sim 0.6 - 1.2$ Gyr, in better agreement with the Sagittarius's pericentric passage. More detailed studies are needed to converge the perturbation times derived from different measurements with the consideration of self-gravity in the future.

There are several factors that can result in spatially varying perturbation times. First, considering a satellite passage, the perturbation is not a delta function, but rather occurs over some finite width of time, depending on the orbital trajectory of the satellite \citep{Darragh-Ford2023}. Second, a subtle difference in the time of peak perturbation for snails in different regions is expected due to the stellar rotation in the disk at the time of interaction \citep{Gandhi2022}, leading to a dipole pattern in the measured interaction times \citep{Darragh-Ford2023}. Third, the self-gravity and the resulting disturbance propagation in the disk can also lead to spatially varying perturbation times, as shown in Fig. 9 of \cite{Asano2026}. Finally, the combined action of the internal (e.g., the bar) and external (e.g., the satellite, DM wake) effects, or multiple perturbations could also lead to radially different perturbation times, or the difference in the two branches of the inner two-armed snails \citep{LinJX2025}.

The comparison of $\theta_{z,0}$ with recent works \citep{Alinder2023, Darragh-Ford2023} is inconvenient due to its large uncertainty and the different phase angles adopted at a certain vertical frequency or radius. \cite{Darragh-Ford2023} adopted the phase angle at $\Omega_z = 45$ {\kms}\,$\rm kpc^{-1}$, while \cite{Alinder2023} utilized the angle in $z-V_z$ space and a phase radius of $r=0.5$ kpc. Nevertheless, their results show variation in the snail phase angle with spatial location and some structures, related to the global rotation of snails. We note that the estimation of $\theta_{z,0}$ is prone to being influenced by $T_{evol}$, and is also degenerate with the potential model.

%%%%%%%%%%%%%%%%%%%%%%%%%%%%%%%%%%%%%%%%%
\subsection{Two-armed spiral and secondary branches}
\label{ssec:dis_twoarm}
%%%%%%%%%%%%%%%%%%%%%%%%%%%%%%%%%%%%%%%%%
Snails are found to widely exist cross the Galactic disk, within a radial range of $R\sim 6-11$ kpc and an azimuthal range of $\phi \sim \pm 20$\textdegree\ \citep{LiZY2021, Hunt2022, Antoja2023}. The outer region of $R\sim 12-17$ kpc and $z$ up to 4 kpc also shows kinematics related to the snails \citep{XuY2020}. Besides, two-armed snails are found at inner radii of $R_g \approx 6.3$ kpc \citep{Hunt2022}, which could be due to multiple origins \citep{Hunt2022, LinJX2025}, a decelerating bar \citep{LiCD2023} or a vertical resonance that breaks the disk adiabaticity \citep{Chiba2025}. \cite{LinJX2025} found that the two branches of the two-armed snails exhibit distinct shapes, with discrepant perturbation times derived for each branch. They explained this observational characteristic by two successive perturbation events separated by a time interval of $\sim 180$ Myr, which occurred at $\sim 320$ Myr and $\sim 500$ Myr ago, respectively. Our explored radial range is currently $R_g \sim 8-11$ kpc, as the outer region is subject to the sample noise due to fewer stars, and the inner region shows the presence of two-armed snails or secondary branches \citep{Antoja2023}. The secondary branch is also reported in \cite{GuoR2024} at $R_g =7.5$ kpc, which may bias the snail shape measurement (at $z \sim 0.7$ kpc) and also the constructed vertical potential profile. The modeling of two-armed snails in the inner region of $R_g \sim 6$ kpc is complicate and possibly subject to the inner non-axisymmetric components (e.g., the Galactic bar). They lead to a smaller perturbation time estimation of $T_{evol} \sim 200-400$ Myr \citep{Frankel2023}. Nevertheless, the clearly detected two-armed snail shapes and their variation with spatial location \citep{Hunt2022, Alinder2024} might be utilized to constrain the properties of the bar. We ignore these regions for the current vertical potential measurement and leave them to more detailed analyses in the future

%%%%%%%%%%%%%%%%%%%%%%%%%%%%%%%%%%%%%%%%%
\subsection{The effect of self-gravity}
\label{ssec:dis_selfg}
%%%%%%%%%%%%%%%%%%%%%%%%%%%%%%%%%%%%%%%%%
We ignore the self-gravity in our method and its validation with the test particle simulation, as its influence is complicated to investigate and requires detailed comparisons with N-body simulations. The self-gravity is thought to gradually change the vertical potential and thus the vertical frequency of stellar oscillation. The gradual variation in the potential should be imprinted in the winding process of the phase snail, and thus in its shape. Comparisons of simulations with and without the self-gravity found that it has important influence on the clarity, tightness, amplitude and duration of the snail \citep{Darling2019, Widrow2023, Asano2026}. The snails in the simulations with the self-gravity are less tightly wound, showing smaller slopes in $\Omega_{z} - \theta_{z}$ space and thus biased to shorter perturbation time. In addition, compared to test particle simulations, the N-body simulation exhibits an initial, coherent vertical oscillation before a clear spiral forms, leading to systematic underestimation of excitation times of $\sim 0.3$ Gyr \citep{Asano2026}.

Nevertheless, the work of \cite{Widmark2022b} provides an estimate of the systematic bias due to the neglecting of self-gravity, which utilized similar method and assumptions to ours. They well reproduced the true vertical potential after considering the constant offset due to the current position of the satellite. They found an unbiased distribution of errors with a standard deviation of 7\% for their 44 data samples, which may suggest that the self-gravity has limited influence to the potential modeling using snails. On the other hand, it is also plausible that the effects of self-gravity have already been imprinted in the inferred potential. In this case, the observed snail-shaped features simply reflect the present-day vertical gravitational potential, rather than the pre-perturbation potential. Moreover, the snail amplitude, i.e., the fraction of stars in the snail feature, is a good indicator of the importance and influence of the self-gravity. More studies on the snail amplitude are helpful to understand the effect of self-gravity \citep[e.g.,][]{Alinder2023, Frankel2023}. In summary, the effect of the self-gravity is complicated and requires a detailed further study in the future.

%%%%%%%%%%%%%%%%%%%%%%%%%%%%%%%%%%%%%%%%%%%%%%%%%%%%%%%%%%%%%%
\section{Conclusions}
\label{sec:con}
%%%%%%%%%%%%%%%%%%%%%%%%%%%%%%%%%%%%%%%%%%%%%%%%%%%%%%%%%%%%%%
We utilized the snail shape curves and the 1D vertical phase-space orbital integration method to measure the vertical potential, based on the linear relation between the vertical frequency, $\Omega_z$, and action angle, $\theta_z$. The snail shape curve is regarded as the ridge line of the snail and extracted from the snail density contrast map, by finding peaks of the density contrast profiles of different wedges in the $z-V_z$ space. Assuming a certain vertical potential, we transformed the $z-V_z$ space snail shape curves into the curves in $\Omega_z - \theta_z$ space, through the 1D orbital integration method used in \cite{Widmark2021a}. The ideal vertical potential would result in a linear relation in $\Omega_z - \theta_z$ space. Utilizing the MCMC technique, we compared the linear performance in $\Omega_z - \theta_z$ space of different potentials according to the linear fitting, to obtain estimations of model parameters, including the potential parameters, the perturbation evolution time, $T_{evol}$, and the initial phase angle, $\theta_{z,0}$.

We verified our method with a test particle simulation built to mimic the impulse approximation by imposing a downward vertical velocity kick. Our method performs well in the test particle simulation. The snail shape curves extracted are well consistent with the density contrast maps and the toy model snails calculated from the input vertical potential. The best-fit potential from MCMC shows a good linear relation in $\Omega_z - \theta_z$ space, with recovered vertical potential profiles in excellent agreement with the input.

We inspected the linear behavior in $\Omega_z - \theta_z$ space of the snail shape curve transformed under arbitrarily scaled input potentials. The globally scaled potentials lead to shifted $\Omega_z - \theta_z$ curves along $\Omega_z$, with obvious wiggles although slightly changed slopes, i.e., $T_{evol}$. In contrast, when the low $z$ and high $z$ regions are scaled oppositely, the slope $T_{evol}$ can be biased significantly, though the $\Omega_z - \theta_z$ curves show no clear wiggles.

We then applied our method to eight $R_{g}$-binned snails within $8 \le R_g \le 11$ kpc from the {\gaia} DR3. We adopted a potential model consisting of a fixed Plummer bulge, a Miyamoto-Nagai disk, and an NFW DM halo, and let $T_{evol}$ and $\theta_{z,0}$ vary with radius. We first tried to measure the vertical potential with the observed snail shape curves only. The $\Omega_z - \theta_z$ curves transformed under the best-fit potential show good linear relations at most radii, especially at inner ones with more points. The vertical potential profiles predicted are also consistent with those from the Milky Way potential model MWPotential2014 \citep{Bovy2015}, when the potential radius, $R_{pot}$, correction is considered for the $R_g$-binned snails. However, the best-fit potential leads to a flatter rotation curve than the observed one from \cite{ZhouY2023}. This discrepancy is alleviated when we combine the radial constraint from the rotation curve with the vertical constraint from the snails. We obtain a roughly similar linear performance in $\Omega_z - \theta_z$ space and slightly steeper vertical potential profiles for this combined model.

Our models of utilizing the snail shape curves only and of combining the constraints from both the phase snails and the rotation curve result in similar constraints on the potential parameters. For the disk parameters, the model using snail shape curves alone predicts a stellar disk mass of $M_d=6.57^{+0.55}_{-0.65} \times 10^{10}{\rm M}_{\odot}$, a scale length of $a_d=4.37^{+0.68}_{-0.46}$ kpc, and a scale height of $b_d=0.285^{+0.047}_{-0.041}$, while the model with the additional constraint from rotation curve predicts: $M_d=7.18^{+1.15}_{-0.63} \times 10^{10}{\rm M}_{\odot}$, $a_d=3.94^{+0.72}_{-0.29}$ kpc, and $b_d=0.306^{+0.014}_{-0.025}$ kpc. Both stellar masses are slightly higher than the broad Gaussian prior adopted ($5.17\pm1.11 \times 10^{10}{\rm M}_{\odot}$). For the disk scale length, $a_d$, considering its difference from the scale length, $R_d$, of an exponential disk, we find our measurements correspond to $R_d \sim 2-2.8$ kpc, using the examples shown in \cite{Smith2015} and \cite{Pouliasis2017}. For the DM halo properties, the model using snails only derives a virial mass of $M_{200}= 0.81^{+0.22}_{-0.23} \times10^{12}\, {\rm M_{\odot}}$, a virial radius of $R_{200}= 193^{+16}_{-20}$ kpc, and a concentration parameter of $c= 11.3^{+2.4}_{-1.9}$, while these parameters for the model combining the rotation curve constraint are $M_{200}= 0.87^{+0.31}_{-0.33} \times10^{12}\, {\rm M_{\odot}}$, $R_{200}= 198^{+21}_{-29}$ kpc, and $c= 13.9^{+11.2}_{-2.6}$. The measurements of $M_{200}$ and $R_{200}$ are both consistent with recent estimations. The larger concentration of the latter model results from the weakly declining rotation curve, with the larger uncertainties due to the discrepancy of mass distributions at large radii ($R> 15$ kpc). The local DM densities of the two models are $\rho_{dm,\odot}= 0.0076^{+0.0019}_{-0.0021} \,{\rm M_{\odot}\,pc^{-3}}$ and $\rho_{dm,\odot}= 0.0097^{+0.0002}_{-0.0009} \,{\rm M_{\odot}\,pc^{-3}}$, for the constraints using snails only and combining the rotation curve, respectively. The high precision of the latter benefits from the combined constraints from the snails and rotation curve in the solar neighborhood, though the posterior distributions of other DM parameters show double peaks. These values are consistent with the studies using global analyses, and are slightly smaller than those of local analyses \citep{deSalas2021}.

The predicted time since the perturbation, $T_{evol}$, in our work is in the range of 400-750 Myr with a median of 564 Myr. Our measurement is consistent with the rough ranges obtained in recent works. The different weakly decreasing trend could be due to the self-consistent potential model derived from the snail shape curves. Obtaining more consistent and convergent measurements of $T_{evol}$ requires one to distinguish the differences in the methods and potential models applied. The initial phase angle, $\theta_{z,0}$, in our work is always close to $\pi$, indicating a perturbation similar to a downward vertical velocity kick.

Our method of measuring the vertical potential utilizing the snail shape curves is efficient and flexible, and could be applied to more radii and azimuthal angles in the future to explore the 3D potential maps and mass distributions. This idea might also be applied to the two-armed snails at inner radii to constrain the properties of the non-axisymmetric component; for example, the Galactic bar. The radial constraint from the observed rotation curve could also help to better constrain the global mass distribution of our Milky Way. Our work demonstrates the wealth information contained in the phase snails and the promising application in determining the Galactic potential.

%%%%%%%%%%%%%%%%%%%%%%%%%%%%%%%%%%%%%%%%%%%%%%%%%%%%%%%%%%%%%%
\begin{acknowledgements}
    We thank the anonymous reviewers for their thoughtful comments.
    We thank Neige Frankel for kindly providing the measurements of perturbation time for comparisons. The research presented here is partially supported by the National Natural Science Foundation of China under grant Nos. 12588202, 12103031, 12233001, 12625301, 12533004, 11773052; by the CAS Project for Young Scientists in Basic Research, grant No. YSBR-062; by the self-initiated research projects of the National Astronomical Observatories of China (NAOC) under grant No. E4ZR0512; by the National Key R\&D Program of China under grant No. 2024YFA1611602; by the China Manned Space Program with grant Nos. CMS-CSST-2025-A08, CMS-CSST-2025-A09 and CMS-CSST-2025-A11; by the ``111'' Project of the Ministry of Education of China under grant No. B20019; and by a Shanghai Natural Science Research Grant (24ZR1491200). RG is supported by Initiative Postdocs Supporting Program (No. BX2021183), funded by China Postdoctoral Science Foundation. This project is supported in part by Office of Science and Technology, Shanghai Municipal Government (grant Nos. 24DX1400100, ZJ2023-ZD-001).

    This project was developed in part at the ``Winding, Unwinding and Rewinding the Gaia Phase Spiral'' workshop hosted by the Lorentz Center in August 2025, Leiden, The Netherlands.  
    This work made use of the Gravity Supercomputer at the Department of Astronomy, Shanghai Jiao Tong University.
    This work has made use of data from the European Space Agency (ESA) mission {\it Gaia} (\url{https://www.cosmos.esa.int/gaia}), processed by the {\it Gaia} Data Processing and Analysis Consortium (DPAC, \url{https://www.cosmos.esa.int/web/gaia/dpac/consortium}). Funding for the DPAC has been provided by national institutions, in particular the institutions participating in the {\it Gaia} Multilateral Agreement.

    This work makes use of the following software packages: {\tt Astropy} \citep{Astropy2013, Astropy2018}, {\tt Numpy} \citep{Numpy2020}, {\tt Scipy} \citep{Scipy2020}, {\tt Matplotlib} \citep{Matplotlib2007}, {\tt Jupyter} \citep{Jupyter2016}, {\tt EMCEE} package \citep{Foreman-Mackey2013}, {\tt AGAMA} \citep{Vasiliev2019}.

\end{acknowledgements}

% Please note that we have included the references below in
% order to compile the document, but we ask you to:

%\begin{thebibliography}{}
  
%\end{thebibliography}{}

\bibliographystyle{aa} % style aa.bst
\bibliography{AA_references}

% %%%%%%%%%%%%%%%%%%%%%%%%%%%%%%%%%%%%%%%%%%%%%%%%%%%%%%%%%%%%%%
% Example below of non-structurated natbib references  
% To use the v8.3 macros with this form of composition of bibliography,
% the option "bibyear" should be added to the command line
% "\documentclass[bibyear]{aa}".
% %%%%%%%%%%%%%%%%%%%%%%%%%%%%%%%%%%%%%%%%%%%%%%%%%%%%%%%%%%%%%%

% \begin{thebibliography}{}

%   \bibitem[1966]{baker} Baker, N. 1966,
%       in Stellar Evolution,
%       ed.\ R. F. Stein,\& A. G. W. Cameron
%       (Plenum, New York) 333

%    \bibitem[1988]{balluch} Balluch, M. 1988,
%       A\&A, 200, 58

%    \bibitem[1980]{cox} Cox, J. P. 1980,
%       Theory of Stellar Pulsation
%       (Princeton University Press, Princeton) 165

%    \bibitem[1969]{cox69} Cox, A. N.,\& Stewart, J. N. 1969,
%       Academia Nauk, Scientific Information 15, 1

%    \bibitem[1980]{mizuno} Mizuno H. 1980,
%       Prog. Theor. Phys., 64, 544
   
%    \bibitem[1987]{tscharnuter} Tscharnuter W. M. 1987,
%       A\&A, 188, 55
  
%    \bibitem[1992]{terlevich} Terlevich, R. 1992, in ASP Conf. Ser. 31,
%       Relationships between Active Galactic Nuclei and Starburst Galaxies,
%       ed. A. V. Filippenko, 13

%    \bibitem[1980a]{yorke80a} Yorke, H. W. 1980a,
%       A\&A, 86, 286

%    \bibitem[1997]{zheng} Zheng, W., Davidsen, A. F., Tytler, D. \& Kriss, G. A.
%       1997, preprint
% \end{thebibliography}

%%%%%%%%%%%%%%%%%%%%%%%%%%%%%%%%%%%%%%%%%%%%%%%%%%%%%%%%%%%%%%%
% Appendices must be placed after   \end{thebibliography}
% They will be placed automatically on a new page.
%%%%%%%%%%%%%%%%%%%%%%%%%%%%%%%%%%%%%%%%%%%%%%%%%%%%%%%%%%%%%%%
\begin{appendix}
%%%%%%%%%%%%%%%%%%%%%%%%%%%%%%%%%%%%%%%%%%%%%%%%%%%%%%%%%%%%%%%
% In the PDF output, floats should be placed
% under their own appendix, not before the title, nor after the
% title of the next appendix.

% In short appendices, onecolumn floats (\figure*
% or \table*) will generate a blank page.
% To prevent this behaviour, a few examples are provided here. 

% In case you have a lot of floating objects for little text and the 
% LaTeX engine moves the floats away from their context, the command
% \FloatBarrier of the “placeins” package will empty the
% float buffer and place all stored floats in the continuity.

% If you still encounter problems with wide floats placement,
% just use the onecolumn environment throughout the appendices.
%%%%%%%%%%%%%%%%%%%%%%%%%%%%%%%%%%%%%%%%%%%%%%%%%%%%%%%%%%%%%%%

%____________________________________________________________
%       Wide floats at the start of an appendix: first method
%-------------------------------------------------------------
% To prevent a blank page after the start of an appendix:
% - Switch to one \onecolumn first
% - Declare the section title
% - Declare the onecolumn float with the parameter [ht!]
% - Revert to \twocolumn at the end of the section
%\onecolumn
\nolinenumbers
%%%%%%%%%%%%%%%%%%%%%%%%%%%%%%%%%%%%%%%%%%%%%%%%%%%%%%%%%%%%%%
\section{Mass and potential models for different components}
\label{sec:app_mfunc}
%%%%%%%%%%%%%%%%%%%%%%%%%%%%%%%%%%%%%%%%%%%%%%%%%%%%%%%%%%%%%%
We summary the potential, mass functions and rotation curves of different components utilized in this work here, including the Plummer sphere bulge \citep{Plummer1911}, the Miyamoto-Nagai disk \citep{MN1975}, the specific DM halo profile constructed in \cite{Irrgang2013} and the NFW DM halo \citep{NFW1996} adopted for our observational data.

The potential and mass functions of a Plummer sphere are described as
%%%%%%%%%%%%%%%
\begin{equation}
\label{eq:bulgeP}
    \Phi_b(r) = -\frac{GM_b}{\sqrt{r^2 + b_b^2}} ,
\end{equation}
%%%%%%%%%%%%%%%
%%%%%%%%%%%%%%%
\begin{equation}
\label{eq:bulgeM}
    \rho_b(r) = \frac{3b_b^2 M_b}{4\pi (r^2 + b_b^2)^{5/2}} ,
\end{equation}
%%%%%%%%%%%%%%%
where $r$ and $R$ are the spherical and cylindrical radius, respectively, and $r=\sqrt{R^2+z^2}$. $M_b$ and $b_b$ are the scale mass and scale radius of the Plummer sphere. G is the gravitational constant. The corresponding rotation curve can thus be written as
%%%%%%%%%%%%%%%
\begin{equation}
\label{eq:bulgeVc}
    V_{c,\,b}^2 = \frac{G M_b r^2}{\left(r^2 + b_b^2\right)^{3/2}} .
\end{equation}
%%%%%%%%%%%%%%%

The potential and mass functions of a Miyamoto-Nagai disk are expressed as
%%%%%%%%%%%%%%%
\begin{equation}
\label{eq:MNdiskP}
    \Phi_d(R,z) = -\frac{GM_d}{\sqrt{R^2 + \left(a_d + \sqrt{z^2 + b_d^2}\right)^2}} ,
\end{equation}
%%%%%%%%%%%%%%%
%%%%%%%%%%%%%%%
\begin{equation}
\label{eq:MNdiskM}
    \rho_d(R, z) = \frac{b_d^2 M_d}{4\pi} \frac{a_d R^2 + \left(a_d + 3\sqrt{R^2 + z^2}\right)\left(a_d + \sqrt{R^2 + z^2}\right)^2}{\left(z^2 + b_d^2\right)^{3/2}\left(R^2 + \left(a_d + \sqrt{z^2 + b_d^2}\right)^2\right)^{5/2}} ,
\end{equation}
%%%%%%%%%%%%%%%
where $M_d$ is the scale mass, $a_d$ and $b_d$ are the scale lengths. The rotation curve of the Miyamoto-Nagai disk is
%%%%%%%%%%%%%%%
\begin{equation}
\label{eq:MNdiskVc}
    V_{c,\,d}^2(R) = \frac{G M_d R^2}{\left[ R^2 + \left( a_d + b_d \right)^2 \right]^{3/2}} .
\end{equation}
%%%%%%%%%%%%%%%

The DM profile adopted in \cite{Irrgang2013} for Model I is chosen to have the asymptotic behavior of $m_h(<r) \propto  r$ for large $r$ motivated by the observed flat rotation curve, and to have a finite halo mass at a cutoff radius $\Lambda$. The potential and mass functions of this special halo profile are given by
%%%%%%%%%%%%%%%
\begin{equation}
\label{eq:IDMP}
    \Phi_h(r) = 
\begin{cases}
\frac{GM_h}{a_h} \left( \frac{1}{(\gamma - 1)} \ln \left( \frac{1 + \left( \frac{r}{a_h} \right)^{\gamma - 1}}{1 + \left( \frac{\Lambda}{a_h} \right)^{\gamma - 1}} \right) - \frac{\left( \frac{\Lambda}{a_h} \right)^{\gamma - 1}}{1 + \left( \frac{\Lambda}{a_h} \right)^{\gamma - 1}} \right) & \text{if } r < \Lambda, \\
-\frac{GM_h}{r} \frac{\left( \frac{\Lambda}{a_h} \right)^{\gamma}}{1 + \left( \frac{\Lambda}{a_h} \right)^{\gamma - 1}}, & \text{otherwise.}
\end{cases}
\end{equation}
%%%%%%%%%%%%%%%
%%%%%%%%%%%%%%%
\begin{equation}
\label{eq:IDMM}
    m_h(<\!r) =
    \begin{cases}
\dfrac{M_{h}\left(\frac{r}{a_{h}}\right)^{\gamma}}{1 + \left(\frac{r}{a_{h}}\right)^{\gamma-1}} & \textrm{if } r < \Lambda, \\
\dfrac{M_{h}\left(\frac{\Lambda}{a_{h}}\right)^{\gamma}}{1 + \left(\frac{\Lambda}{a_{h}}\right)^{\gamma-1}} = \textrm{constant}, & \textrm{otherwise.}
   \end{cases}
\end{equation}
%%%%%%%%%%%%%%%
Here, $M_h$ and $a_h$ are again the scale mass and scale length, respectively. In \cite{Irrgang2013}, $\gamma$ is fixed as $\gamma=2$, and the cutoff radius $\Lambda$ is constrained as $\Lambda \leq 200$ kpc and has a best-fit value of $\Lambda = 200$ kpc. The rotation curve under conditions of $\gamma=2$ and $r< \Lambda$ is simplified as 
%%%%%%%%%%%%%%%
\begin{equation}
\label{eq:IDMVc}
    V_{c,\,h}^2(r) = \frac{G M_h r}{a_h(a_h + r) }.
\end{equation}
%%%%%%%%%%%%%%%

For the observational data, we apply the NFW DM halo proposed by \cite{NFW1996}. Its density, potential and rotation curve are given by
%%%%%%%%%%%%%%%
\begin{equation}
\label{eq:NFWM}
    \rho_{\mathrm{NFW}}(r) = \frac{\rho_0}{ (r/r_s)(1+r/r_s)^2 } ,
\end{equation}
%%%%%%%%%%%%%%%
%%%%%%%%%%%%%%%
\begin{equation}
\label{eq:NFWP}
    \Phi_{\mathrm{NFW}}(r) = -\frac{4\pi G\rho_0 r_s^3}{r} \ln(1+r/r_s),
\end{equation}
%%%%%%%%%%%%%%%
%%%%%%%%%%%%%%%
\begin{equation}
\label{eq:NFWVc}
    V_{\mathrm{c,\,NFW}}^2(r) = \frac{4\pi G\rho_0 r_s^3}{r} \left[ \ln(1+r/r_s) - \frac{r/r_s}{1+r/r_s} \right] ,
\end{equation}
%%%%%%%%%%%%%%%
where $r_s$ is the scale radius (same as $a_h$ in the main text), and $\rho_0$ is the characteristic DM density. We also adopt a scale mass parameter $M_h= 4\pi \rho_0 r_s^3$ for convenience.

A DM halo is usually described by another set of parameters: the virial radius $r_{200}$, the virial mass $M_{200}$ and the concentration parameter $c= r_{200}/r_s$. $r_{200}$ is defined the radius where the average density inside is 200 times the critical density $\rho_{crit}$ with 
%%%%%%%%%%%%%%%
\begin{equation}
\label{eq:rho_c}
    \rho_{crit}= \frac{3H_0^2}{8\pi G} .
\end{equation}
%%%%%%%%%%%%%%%
In this work, we adopt $\Omega_m= 0.28$ and $H_0= 69.7$ {\kms}\,Mpc$^{-1}$ from \cite{Hinshaw2013} to calculate the critical density $\rho_{crit}$. Thus, the viral mass can be expressed as 
%%%%%%%%%%%%%%%
\begin{equation}
\label{eq:mvir}
    M_{200}= \frac{4\pi}{3} \times 200\rho_{crit} \times R_{200}^3 
           = M_h \times \left[ \ln{(1+c)} - \frac{c}{1+c} \right] .
\end{equation}
%%%%%%%%%%%%%%%
The characteristic DM density $\rho_0$ can also be written as a function of concentration $c$:
%%%%%%%%%%%%%%%
\begin{equation}
\label{eq:rho_0c}
    \rho_0= \frac{200}{3} \frac{c^3}{\ln(1+c) - c/(1+c)} \times \rho_{crit} .
\end{equation}
%%%%%%%%%%%%%%%

%%%%%%%%%%%%%%%%%%%%%%%%%%%%%%%%%%%%%%%%%
\section{Variation in the \texorpdfstring{$\Omega_{z} - \theta_{z}$}{Omegaz - thetaz} profile with scaled input potential}
\label{sec:scaled_pot}
%%%%%%%%%%%%%%%%%%%%%%%%%%%%%%%%%%%%%%%%%
Ideally, the well recovered vertical potential would result in a good linear relation after transforming the phase snail into $\Omega_{z} - \theta_{z}$ space, as shown in Fig. \ref{fig:simu_Oz_thetaz}. Here we investigate the variation in the curve in $\Omega_{z} - \theta_{z}$ space with the vertical potential deviating from the input one, to help understand uncertainties of the potential constrained by the snail shape. We utilize the snail at $R_g =8$ kpc and the arbitrarily scaled input vertical potential from Irrgang13I to convert the snail into $\Omega_{z} - \theta_{z}$ space. These tests are just pure mathematical tests, as the potential relies on the mass model and cannot be scaled casually in observation. We mainly want to investigate the impact on the results under a wrong potential, which can in turn justify the accuracy of the recovered potential.

\begin{figure*}[htbp]
\centering
\includegraphics[width=0.9\textwidth]{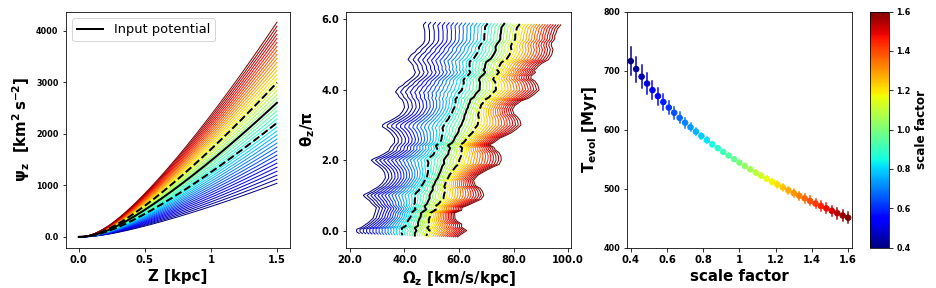}
\caption{
\label{fig:scale_bulk}Globally scaled vertical potential (\textit{left}), the corresponding $\Omega_{z} - \theta_{z}$ curves (\textit{middle}) transformed from the same snail shape curve and their slopes $T_{evol}$ as a function of the scale factor (\textit{right}). The color bar indicates the scale factor ranging from 0.4 (blue) to 1.6 (red). The solid black lines in the left two panels are the input potential and its $\Omega_{z} - \theta_{z}$ curve, while the dashed lines are those with a potential deviation of 15\%. The $\Omega_{z} - \theta_{z}$ curve shows a clear zigzag pattern when the potential is significantly scaled.}
\end{figure*}

We first scale the input potential profile globally by a factor of $\alpha$ ranging from 0.4 to 1.6. The $\Omega_{z} - \theta_{z}$ profiles transformed from the same snail shape curve under these different potential profiles are shown in the middle panel of Fig. \ref{fig:scale_bulk}, with the relation between scale factor $\alpha$ and model predicted $T_{evol}$ illustrated in the right panel. The $\Omega_{z} - \theta_{z}$ curves are still close to straight lines when the scaled potentials are within $\sim$15\% of the original potential profile (dashed black lines). However, when the scaled potential deviates further away from the input, the $\Omega_{z} - \theta_{z}$ curves display a clear zigzag pattern, as shown by the blue and red lines. Nevertheless, these curves have roughly similar slopes, i.e., similar $T_{evol}$ but with different uncertainties. $T_{evol}$ measured from linear fitting to these curves ranges from about 450-700 Myr, showing a negative relation with the scale factor. The small influence on $T_{evol}$ results from the fact that this scaling method globally changes the second-order derivative (similar to $\Omega_{z}$) of the vertical potential, leaving its vertical variation less changed. Thus the $\Omega_{z} - \theta_{z}$ curves shift horizontally along $\Omega_{z}$ axis with similar slopes.

%%%%%%%%%%%%%%%%%%%%%%%%%%%%%%%%%%%%%%%%%
%%%%%%%%%%%%%%%%%%%%%%%%%%%%%%%%%%%%%%%%%

\begin{figure*}[htbp]
\centering
\includegraphics[width=0.7\textwidth]{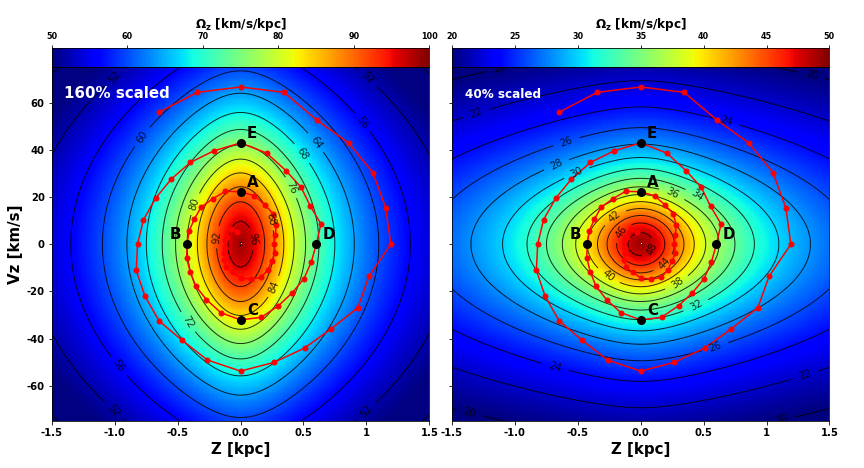}
\caption{
\label{fig:scale_cont}Same snail shape curve superimposed on the vertical frequency maps derived under two extremely scaled potentials. The left and right panels correspond to the 160\% and 40\% scaled potential profiles, respectively. The color maps and contours indicate the vertical frequency $\Omega_z$, and the red curves are the same snail shape curve at $R_{g}=8$ kpc. Five points (A, B, C, D, E) indicated by the black dots are marked to help understand the zigzag pattern of $\Omega_{z} - \theta_{z}$ curves. The $\Omega_z$ contours in the left panel are stretched in the $V_z$ direction, while they are squashed in the right panel. The deformation of the $\Omega_z$ map changes the monotonic variation in $\Omega_z$ along the snail shape curve, resulting in the zigzag pattern of $\Omega_{z} - \theta_{z}$ curves.}
\end{figure*}

To understand the zigzag pattern of $\Omega_{z} - \theta_{z}$ curves in the middle panel of Fig. \ref{fig:scale_bulk}, we overplot the snail shape curve in the $z-V_{z}$ space color coded by the vertical frequency ($\Omega_{z}$) of the potentials with scale factors of 160\% and 40\% (representing two extreme cases). Results are shown in Fig. \ref{fig:scale_cont}. The frequency contour of the 160\% scaled potential shown in the left panel is obviously stretched along the $V_z$ axis, while that of the 40\% scaled potential displayed in the right panel is more elongated in the $z$ direction. The deformation of the frequency contour, or equally the $E_z$ contour, is similar to the variation in the snail shape observed at different radii \citep{LiZY2021}: snails at inner radii are more elongated along the $V_z$ axis due to the steeper vertical potential, while those at outer radii are more stretched in the $z$ direction due to the shallower potential.

In principle, from the central to the outer region of $z-V_{z}$ space, the points on the snail shape curve should have increasing vertical energy and decreasing vertical frequency under the input potential. However, under the scaled potential, the deformation of the frequency contour changes this trend. For example, we marked five points (A, B, C, D, and E) of the input snail shape curve with black points in Fig. \ref{fig:scale_cont}, which are the intersections between the snail and $z/V_{z}$ axes. Ideally, the frequencies of these points should follow: $\Omega_{z, A} > \Omega_{z, B} > \Omega_{z, C} > \Omega_{z, D} > \Omega_{z, E}$. However, for the 160\% scaled potential in the left panel of Fig. \ref{fig:scale_cont}, the stretch of the frequency contour along $V_z$ axis results in $\Omega_{z, B} < \Omega_{z, C}$ and $\Omega_{z, D} < \Omega_{z, E}$. On the other hand, for the 40\% scaled potential in the right panel, the squash of the frequency contour along $V_z$ axis leads to $\Omega_{z, A} < \Omega_{z, B}$ and $\Omega_{z, C} < \Omega_{z, D}$. Therefore, under a globally scaled potential, the derived $\Omega_{z}$ along the snail shape curve would fluctuate, rather than decreasing monotonically, leading to the zigzag pattern of the $\Omega_{z} - \theta_{z}$ curve in the middle panel of Fig. \ref{fig:scale_bulk}.

%%%%%%%%%%%%%%%%%%%%%%%%%%%%%%%%%%%%%%%%%
%%%%%%%%%%%%%%%%%%%%%%%%%%%%%%%%%%%%%%%%%

%%%%%%%%%%%%%%%%%%%%%%%%%%%%%%%%%%%%%%%%%
\begin{figure*}[htbp]
\centering
\includegraphics[width=0.9\textwidth]{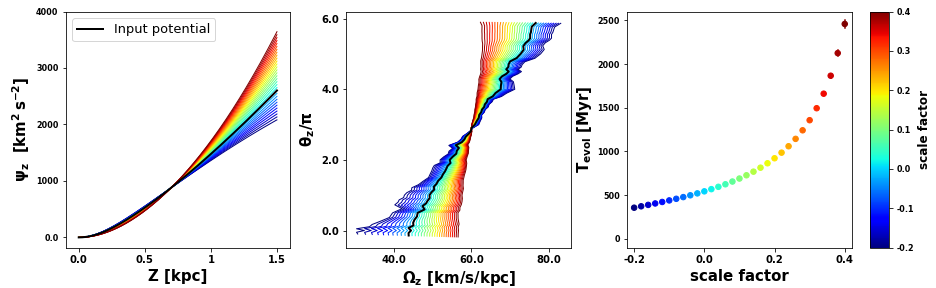}
\caption{
\label{fig:scale_up}Same as Fig. \ref{fig:scale_bulk} but for a different potential scaling function, which is a second-order polynomial of $z$. The potential at $z= 0.7$ kpc is fixed, while the potential at $z= 0$ kpc and $z= 1.5$ kpc are scaled by factors of $1-\alpha$ and $1+\alpha$, respectively. The color-bar stands for $\alpha$ values ranging from $-0.2$ (blue) to 0.4 (red).}
\end{figure*}
%%%%%%%%%%%%%%%%%%%%%%%%%%%%%%%%%%%%%%%%%

We further investigate the influence of a different potential scaling function on the $\Omega_{z} - \theta_{z}$ curves and the derived $T_{evol}$. We utilize a second-order polynomial function of $z$ to scale the vertical potential profile, where the potential at $z= 0.7$ kpc is kept unchanged, while the potential at $z= 0$ kpc and $z= 1.5$ kpc are scaled by factors of $1-\alpha$ and $1+\alpha$, respectively. Specifically, the scale factor varies with $z$ as a function of $-\frac{5}{42}\alpha z^2 + \frac{127}{84}\alpha z + 1-\alpha$. This complicated potential scaling is adopted to scale the potential in low and high $z$ regions conversely, which will alters the vertical frequency profile significantly compared to the input. Figure \ref{fig:scale_up} shows the results with the scale factor $\alpha$ ranging from $-0.2$ to 0.4. Different from the global scaling function, the $\Omega_{z} - \theta_{z}$ curves display roughly linear relations without significant zigzag pattern, as shown in the middle panel of Fig. \ref{fig:scale_up}. As this kind of potential scaling perturbs the vertical profile of the second-order derivative of the potential (i.e., $\Omega_z$) significantly, it thus leads to dramatically changed slopes of $\Omega_{z} - \theta_{z}$ curves and biased $T_{evol}$ in some cases, as shown in the right two panels of Fig. \ref{fig:scale_up}. The red lines, when the high $z$ region is scaled up and the low $z$ region is scaled down, imply the $T_{evol}$ estimation up to 2500 Myr, significantly deviated from the true value with a 5 times overestimation. In the opposite cases, i.e., the blue lines, the slopes are weakly biased to lower values. As these cases show roughly linear performance in $\Omega_{z} - \theta_{z}$ space, our method may not well distinguish them from the true potential, resulting in large uncertainty in the recovered potential and strong degeneracy between the potential parameters and the perturbation time, $T_{evol}$.

% If text follows like so 
%\lipsum[1-2]
% it is easier to finish the page in onecolumn and revert to
% twocolumn when starting the next page if needed.}

%\FloatBarrier %\usepackage{placeins}
%\twocolumn
%____________________________________________________________
%       Wide floats at the start of an appendix: second method
%-------------------------------------------------------------
% To prevent a blank page, a second method is:
% - Declare the onecolumn float *
% - Declare the section under the float
% However, this method should be reserved to appendices
% containing only onecolumn tables or figures.

%\FloatBarrier %\usepackage{placeins}
\clearpage

\end{appendix}
\end{document}